\documentclass{aa}  

\usepackage{graphicx}
\usepackage{color}

\usepackage{txfonts}
\usepackage{hyperref}
\newcommand{\gaia}{\textit{Gaia}}
\definecolor{dgreen}{rgb}{0.1, 0.53, 0.22}
 
\usepackage{lineno}
\begin{document} 
   \title{Cepheid Metallicity in the Leavitt Law (C- MetaLL) survey:} 
   \subtitle{VIII: Spectroscopic detection of rare earth dysprosium, erbium, lutetium, and thorium in classical Cepheids}
\titlerunning{C-MetaLL Survey VIII}
   \author{E. Trentin\inst{1}\thanks{E-mail: erasmo.trentin@inaf.it}
          \and
          G. Catanzaro\inst{2}
          \and 
          V. Ripepi\inst{1}
          \and 
          E. Luongo\inst{1,3}
          \and 
          M. Marconi\inst{1}
          \and 
          I. Musella\inst{1}
          \and 
          F. Cusano\inst{4}
          \and 
          J. Storm\inst{5}
          \and 
          A. Bhardwaj\inst{6}
          \and
          G. De Somma\inst{1,7}
          \and 
          S. Leccia\inst{1}
          \and 
          T. Sicignano\inst{1,7,8,9}
          \and 
          R. Molinaro\inst{1}
          \and 
          V. Testa\inst{10}
          }

\institute{INAF-Osservatorio Astronomico di Capodimonte, Salita Moiariello 16, 80131, Naples, Italy
\and
INAF-Osservatorio Astrofisico di Catania, Via S.Sofia 78, 95123, Catania, Italy 
\and
Università di Salerno, Dipartimento di Fisica “E.R. Caianiello”, Via Giovanni Paolo II 132, 84084 Fisciano (SA), Italy.
\and
INAF-Osservatorio di Astrofisica e Scienza dello Spazio, Via Gobetti 93/3, I-40129 Bologna, Italy 
\and
Leibniz-Institut f\"{u}r Astrophysik Potsdam (AIP), An der Sternwarte 16, D-14482 Potsdam, Germany
\and
Inter-University Center for Astronomy and Astrophysics (IUCAA), Post Bag 4, Ganeshkhind, Pune 411 007, India
\and
Istituto Nazionale di Fisica Nucleare (INFN)-Sez. di Napoli, Via Cinthia, 80126 Napoli, Italy
\and
European Southern Observatory, Karl-Schwarzschild-Strasse 2, 85748 Garching bei München, Germany 
   \and 
Scuola Superiore Meridionale, Largo San Marcellino10 I-80138 Napoli, Italy
\and
INAF – Osservatorio Astronomico di Roma, via Frascati 33, I-00078 Monte Porzio Catone, Italy 
}
        
   \date{Received September 15, 1996; accepted January 9, 2026}

 
  \abstract
   {Classical Cepheids (DCEPs) are among the most important distance calibrators thanks to the correlation between their period and luminosity (PL relation), and play a crucial role in the calibration as the first rung of the extragalactic distance ladder. Given their typical age, they also constitute an optimal tracer of the young population in the Galactic disc.}
   {We aim to increase the number of available DCEPs with high-resolution spectroscopic metallicities, study the galactocentric radial gradients of several chemical elements, and analyse the spatial distribution of the Galactic young population of stars in the Milky Way disc.}
   {We performed a complete spectroscopical analysis of 136 spectra obtained from three different high-resolution spectrographs, for a total of 60 DCEPs. More than half have pulsational periods longer than 15 days, up to 70 days, doubling the number of stars in our sample with $P>15d$. We derived radial velocities, atmospheric parameters, and chemical abundances for up to 33 different species.}
   {We present an updated list of trusted spectroscopic lines for the detection and estimation of chemical abundances. We used this new set to revisit the abundances already published in the context of the C-MetaLL (Cepheids-Metallicity in the Leavitt Law) survey and increase the number of available chemical species. For the first time (to our knowledge), we present the estimation of abundances for Cepheids for dysprosium (Dy, $ Z=$66), as well as a systematic estimation of erbium (Er, $Z=$68), lutetium (Lu, $Z=$71), and thorium (Th, $Z = $90) abundances.}
   {We calculated a galactic radial gradient for $[Fe/H]$ with a slope of $-0.064\pm0.002$ dex $kpc^{-1}$, in good agreement with recent literature estimation. The other elements also exhibit a clear negative radial trend, with this effect diminishing and eventually disappearing for heavier neutron-capture elements. Depending on the proposed spiral arms model present in several literature sources, our most external stars agree on tracing either the Perseus, the Norma-Outer, or both the Outer and the association Outer-Scutum-Centaurus (OSC) arms.}


   \maketitle
%

\section{Introduction}
Classical Cepheids (DCEPs) have been proven to play a fundamental role as standard candles, thanks to their relation between pulsational period and luminosity \citep[PL relation, ][]{Leavitt1912}. The calibration of this relation, or its reddening-free version, the period-Wesenheit \citep[PW, ][]{Madore1982} relation, through independent distance estimation using geometric methods such as trigonometric parallaxes, eclipsing binaries, and water masers, allows for the determination of distances of faraway objects, such as galaxies containing both DCEPs and type Ia supernovae.
This further allows for the measurement of distances of more distant galaxies in the steady Hubble flow and eventually of one of the most important astrophysical parameters, the Hubble constant ($H_0$). This parameter conceals within its value information on both the acceleration of the Universe and its age, making its determination with at least a 1\% precision of extreme importance in modern astrophysics. Nowadays, several estimations are offered in the literature, exploring different methods. While the Planck mission, with the study of the cosmic microwave background under the assumption of the $\Lambda$ cold dark matter ($\Lambda$CDM) cosmological model obtained $H_0$ = 67.4$\pm$0.5 km $s^{-1}$ $Mpc^{-1}$ \citep{Planck2020}, the Supernovae, H0, for the Equation of State of Dark energy (SH0ES) project, using the cosmic distance ladder, estimated $H_0$ = 73.01$\pm$0.99 km $s^{-1}$ $Mpc^{-1}$ \citep{Riess2022a}. No solution has yet been reached for this 5 $\sigma$ discrepancy, but other empirical methods were proposed to analyse possible systematics in the calibration of the cosmic distance ladder, based either on replacing DCEPs with other astronomical objects, such as asymptotic giant branch stars \citep{lee2025,siyang2025}, the tip of the red giant branch \citep{hoyt2025}, and Miras \citep{Bhardwaj2025}, or on replacing both the DCEPs and type Ia supernovae with red giant branch stars and surface brightness fluctuations\citep{anand2025}, respectively.
The metallicity dependence of PL/PW relations (also called PLZ/PWZ relations) may play a crucial role as a source of uncertainty \citep[e.g. see recent discussion in ][]{Riess2021,Bhardwaj2023,madore2025,breuval2025}.
In this context, we initiated a project named Cepheid - Metallicity in the Leavitt Law (C-MetaLL, see \citet{Ripepi2021a}), whose main aim is to collect high-resolution galactic DCEP spectra and build a sample of more than 400 targets with spectroscopic metallicities estimated homogeneously \citep{Ripepi2021a,Trentin2023a,trentin2024}. High-resolution spectroscopy is also being complemented with homogeneous time-series photometry at multiple wavelengths in our survey (e.g. \citet{Bhardwaj2024}). In particular, it is essential to obtain a dataset of targets well distributed in both metallicities and the period range to cover all the phase space for the calibration of the PLZ relations.

\begin{figure}
	\includegraphics[width=9cm]{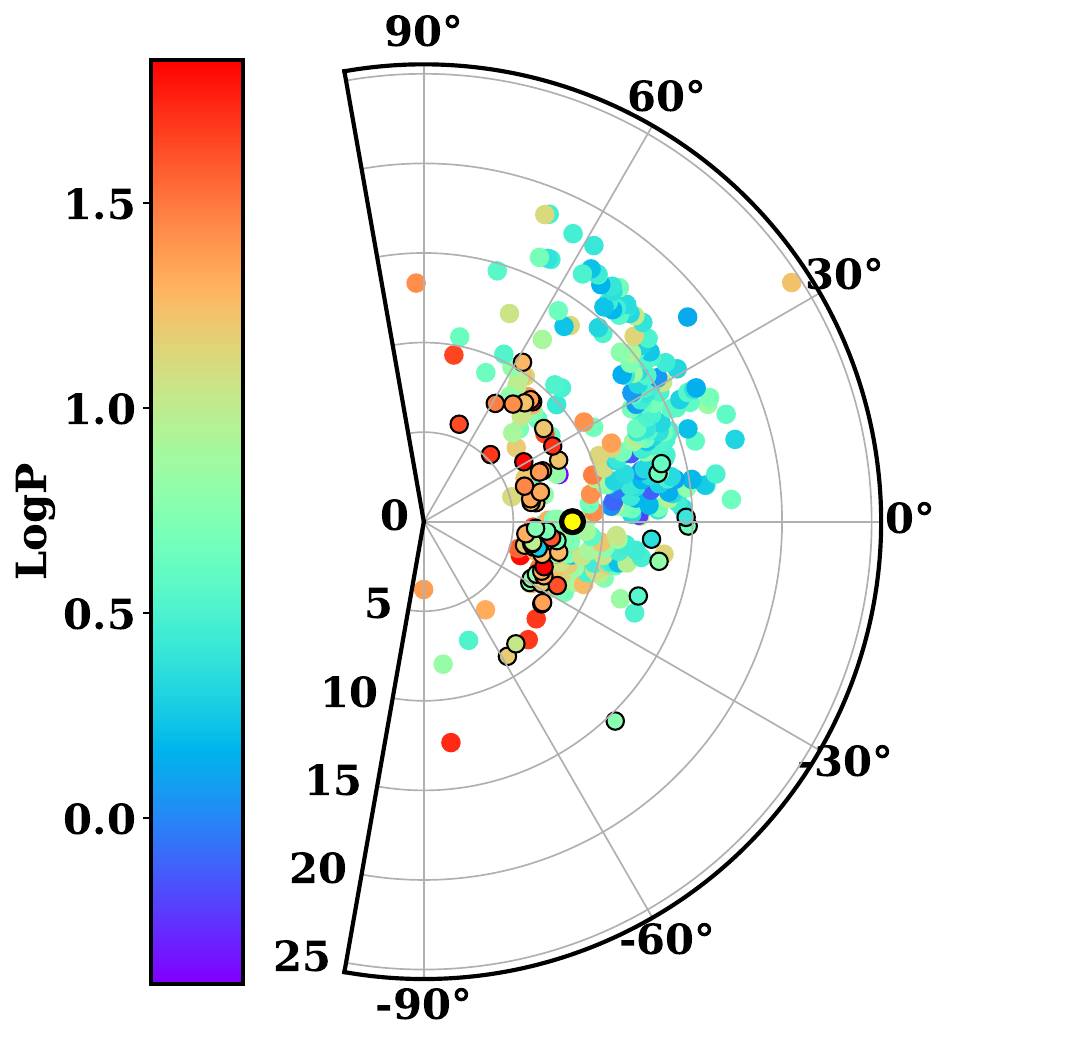}
    \caption{Distribution of the C-MetaLL sample along the Galactic disc. The points are colour-coded according to the pulsator's period. Targets presented in this work are highlighted with black borders. The position of the sun is shown with a yellow-black circle.}
    \label{fig:mapTargets}
\end{figure}

Our project has been very successful in increasing the number of DCEPs with low metallicity \citep[see ][]{Trentin2023a,trentin2024}, but there is still a need to increase the number of stars with periods longer than 15 days.
Beyond their role in the construction of the cosmic distance ladder, DCEPs are useful tracers of young population stars. In this role, they have also been used in the Milky Way (MW) \citep[see e.g.][for a recent review of the role of DCEPs in the Galaxy]{bono2024}, and several recent studies have focused on the use of DCEPs to study the chemical distribution and the spiral structure of the MW \citep[][and reference therein]{Luck2011,Genovali2014,Luck2018,lemasle2018milky,Skowron2019,Minniti2020,Ripepi2021a,Lemasle2022,Kovtyukh2022,daSilva2023,Trentin2023a,Matsunaga2023,trentin2024,drimmel2025,desomma2025}. The possibility of estimating chemical abundances for several elements, ranging from light elements such as carbon and oxygen up to heavy elements produced by neutron capture events such as yttrium (Y), barium (Ba), and europium (Eu), offers a unique opportunity to study the origin of these elements and discriminate different nucleosynthesis channels \citep[see e.g. ][ and reference therein]{kobayashi2020}.
This is the eighth paper within the C-MetaLL project and the fourth in which we present new observations and spectroscopic data to enlarge the sample of DCEPs with homogeneous high-resolution metallicities \citep[see e also][]{Ripepi2021a,Trentin2023a,trentin2024}.
The structure of the paper is as follows: in Sect.~\ref{sec:observations}, we describe the sample of new DCEPs; in Sect.~\ref{sec:spec}, we describe the procedure to estimate both the atmospheric parameters and abundances; results are presented and discussed in Sect.~\ref{sec:results} and Sect.~\ref{sec:discussion}, respectively. We give our conclusion in Sect.~\ref{sec:conclusion} 

\section{Observations and data reduction}\label{sec:observations}

In the context of the C-MetaLL survey, the observations were collected using three different instruments:

\begin{itemize} 
    \item  
    A sample of 23 stars (29 spectra in total) were observed at the European Southern Observatory (ESO) during period P112 (2023-2024) under proposal 112.25NA. We used the Ultraviolet and Visual Echelle Spectrograph \citep[UVES,][]{UVES_Dekker2000}\footnote{\url{https://www.eso.org/sci/facilities/paranal/instruments/uves.html}}, attached to Unit Telescope 2 (UT2) of the Very Large Telescope (VLT), placed at Paranal (Chile). The red arm was used, equipped with the grism CD\#3, covering the wavelength interval 4760--6840 {\AA}, and with the central wavelength at 5800 {\AA}. The 1 arcsec slit, which provides a dispersion of R$\sim$47,000 (sampling 2~px), was selected for all the targets. The number of epochs obtained with this instrument is typically 1-2.
    
    \item 
    A sample of 15 stars (for a total of 44 spectra) was observed with the High Accuracy Radial velocity Planet Searcher for the Northern hemisphere \citep[HARPS-N,][]{HARPS_Mayor2003,HARPS_Cosentino2012} \footnote{https://www.tng.iac.es/instruments/harps/}), at the 3.5m Telescopio Nazionale Galileo (TNG), during period 47 (proposal A47TAC\_18, July-August 2024). HARPS-N features an echelle spectrograph covering the wavelength range between 3830 and 6930 {\AA}, with a spectral resolution of R=115,000 (sampling 3.3~px). Three or more epochs were obtained with HARPS-N for each object.

    \item A sample of 22 stars (for a total of 63 spectra) was observed with the Potsdam Echelle Polarimetric and Spectroscopic Instrument \citep[PEPSI,][]{Strassmeier2015}, the fibre-feed high-resolution optical echelle spectrograph for the Large Binocular Telescope (LBT). The observations were carried out in two periods (proposals IT-2019B-014 and IT-2021\_2022\_24) in 2019-2020 and 2021-2022.     
    The wavelength coverage is between 4750 and 7430 {\AA}, with a spectral resolution of R=50000. Up to four epochs were obtained for each star with this instrument.

\end{itemize}

Reduction of all the spectra, which included bias subtraction, spectrum extraction, flat-fielding, and wavelength calibration, was done by automatic pipelines provided by the three instrument teams, so that we downloaded the science-ready one-dimensional spectra. For more details on the data reduction of HARPS-N, UVES spectra, see \citet{Ripepi2021a} and \citet{Trentin2023a}, respectively. The PEPSI data were reduced using the PEPSI ‘Software for Stellar Spectroscopy’ \citep[SDS4PEPSI][]{Strassmeier2018}.

Table~\ref{tab:programStars} lists the main properties of the target stars (RA, Dec, Period, pulsational mode, \gaia~photometry and galactocentric polar coordinates), while the heliocentric Julian day (HJD) of observation, the exposure time and the signal-to-noise ratio (S/N) for each spectrum are shown in Table~\ref{tab:logObservations}.

The procedure to estimate some of the listed quantities in Table~\ref{tab:programStars} is described in \citet{gaia2023} and \citet{Ripepi2022a}. In summary, we calculated both the apparent Wesenheit magnitude $W=G-1.90\times(G_{BP}-G_{RP}$) using the \gaia~photometry \citep{Ripepi2019} and the absolute W magnitude from the PWZ relation $W= (-5.988 \pm 0.018) -(3.176 \pm 0.044)(\log P -1) -(0.520 \pm 0.09)[Fe/H]$ \citep{Ripepi2022a}. This way, it was possible to obtain the distance modulus, and so the distance. We converted the coordinates into the Galactocentric system, adopting $R_0 = 8277 \pm 9$(stat) $\pm 30$(sys) pc as the distance of the Sun from the Galactic centre \citep{GRAVITY2021}.

Some stars were also observed previously within the C-MetaLL project, and their abundances published elsewhere. In more detail: 
\begin{itemize}
    \item 
    TX Sct, ASAS-SN-J193606.44+215011.4, and ET Vul were already observed with HARPS-N, and had abundances published in \citet{Ripepi2021a}   and \citet{trentin2024}.
    \item                                           
 V459 Sct, V480 Aql, OGLE-GD-CEP-0604, and OGLE-GD-CEP-0841 were observed with UVES during ESO P105 (the first two), P109 and P110, respectively. Abundances were published in \citet{Trentin2023a} and \citet{trentin2024}. 
\end{itemize}

These stars are treated similarly to other stars with multiple observations (see Sect.\ref{sec:abundances}) and, when observed with multiple instruments, represent a further test for confirming that there is no bias in the abundances.

Two stars, Gaia DR2 4638315266636230400 and Gaia DR2 4644478166748030976 (also known as OGLE SMC-CEP-4986 and OGLE SMC-CEP-4953, respectively), are not MW stars. Their spectral analysis will be included in this work, but the targets will be excluded from the results unless specifically mentioned (see Sect.~\ref{sec:results}). We note that 38 objects have periods longer than 15d, for a maximum of 70 days. These objects are particularly useful, as they are rarer than the short-period ones and crucial to defining the PL/PW relations over a large range of periods. The distribution of the whole sample of stars included in the C-MetaLL project is displayed in Fig.~\ref{fig:mapTargets}.

\begin{figure}
	\includegraphics[width=9cm]{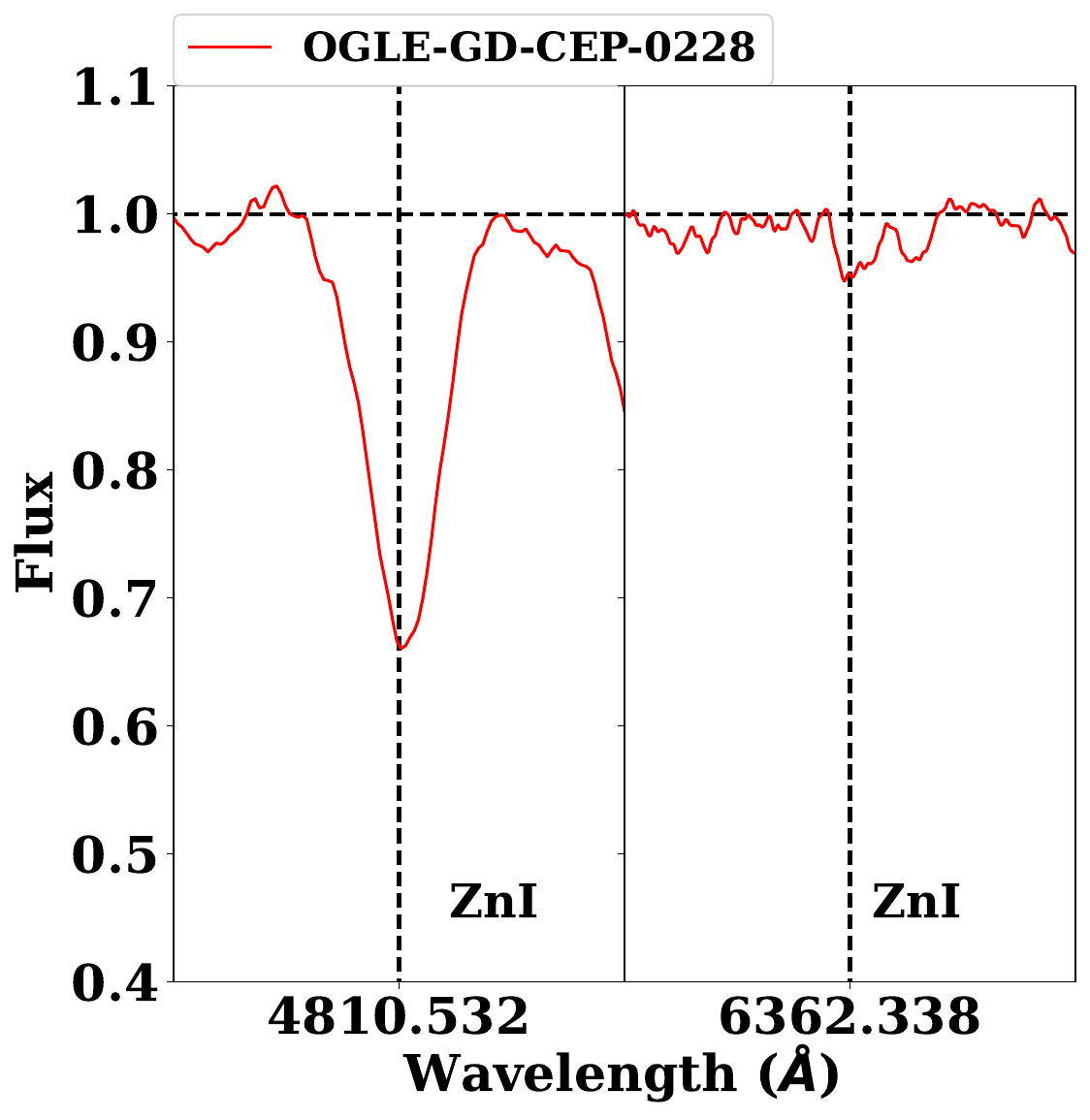}
    \caption{Comparison of two neutral Zinc lines at $\lambda$\, = 4810.532 (left panel) and $\lambda$\, = 6362.338 (right panel){\AA}, for the same star (OGLE-GD-CEP-0228). The two spectral lines are highlighted with vertical dashed black lines, while the horizontal dashed black line defines the continuum (set at 1).}
    \label{fig:Zinc}
\end{figure}

\begin{figure*}
	\includegraphics[width=18cm]{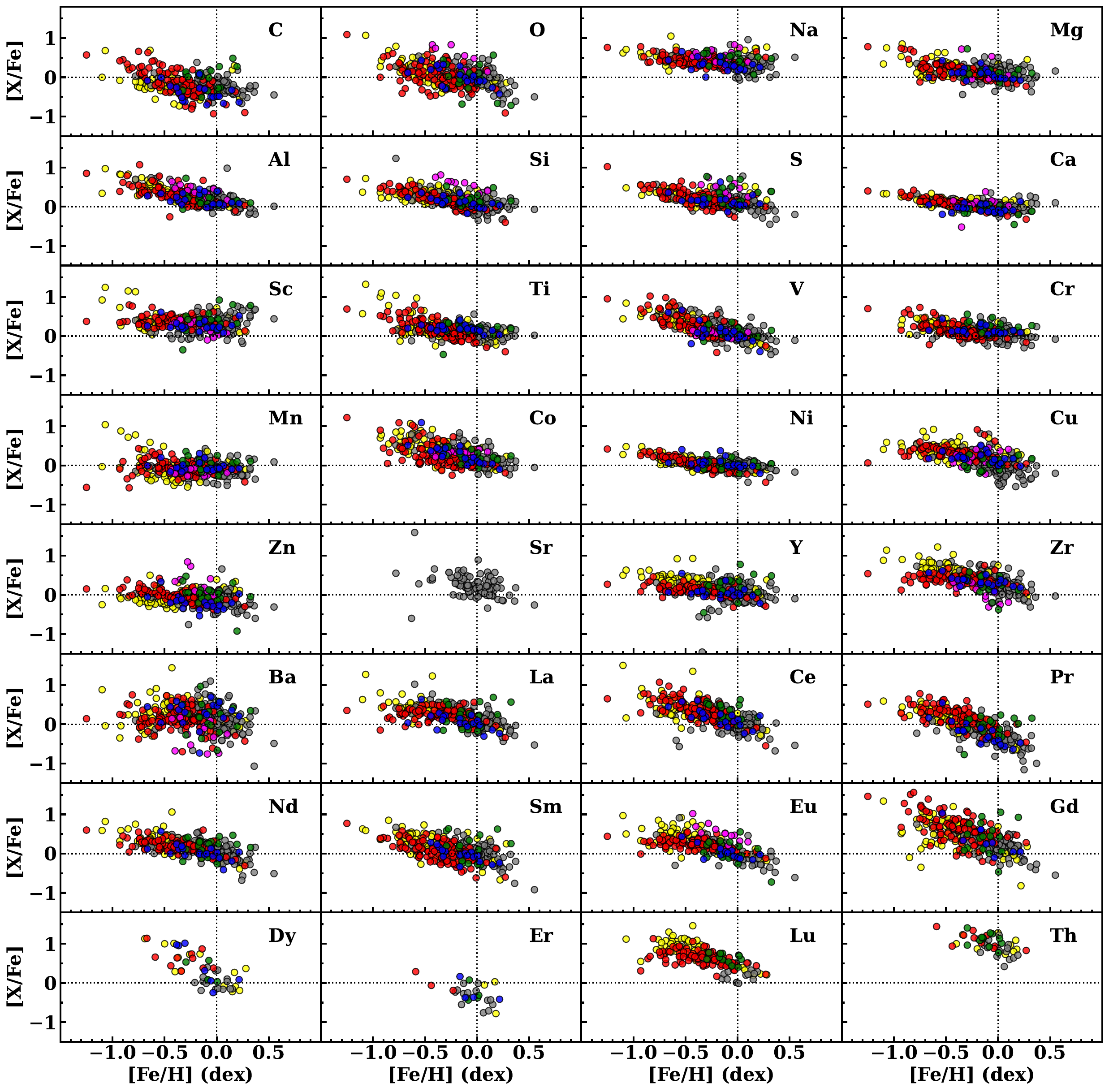}
    \caption{Chemical elements in the form of [X/Fe] plotted against iron ([Fe/H]). Grey points represent stars already published in \citet{Ripepi2021a,Trentin2023a,trentin2024}. Stars observed with UVES and published in \citet{Trentin2023a,trentin2024} are coloured in yellow and red, respectively. New stars presented in this paper are coloured in blue, magenta, and green for PEPSI, HARPS, and UVES observations, respectively (see also the legend in \ref{fig:iron_grad}). Vertical and horizontal dashed grey lines highlight the solar abundance position in the plot.}
    \label{fig:ab_vs_iron}
\end{figure*}

\begin{figure*}
	\includegraphics[width=18cm]{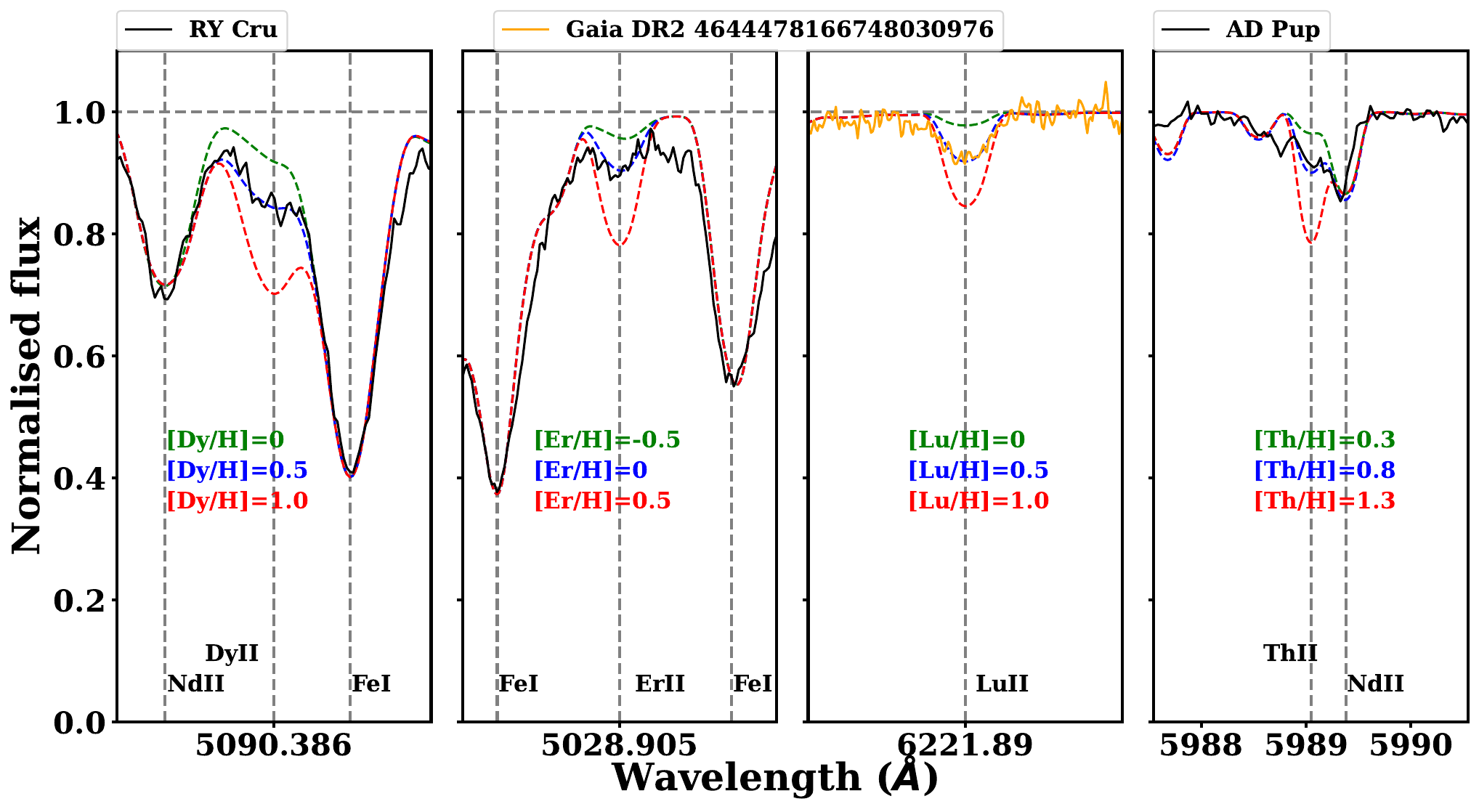}
    \caption{Observed spectrum of RY Cru (solid black line, left and centre panel) and Gaia DR2 4644478166748030976 (solid orange line, right panel) in the region of DyII (5090.386 {\AA}, left panel), ErII (5028.905 {\AA}, centre panel) and LuII (6221.89 {\AA}, right panel) spectral lines. The position of these and other visible lines is reported and highlighted with vertical dashed grey lines. Synthetic spectra computed for the best estimated abundance are reported in blue, while the same spectra with [X/H]=$\pm$0.5 dex are plotted in red and green, respectively.}
    \label{fig:r_elements}
\end{figure*}

\begin{figure*}
	\includegraphics[width=18cm]{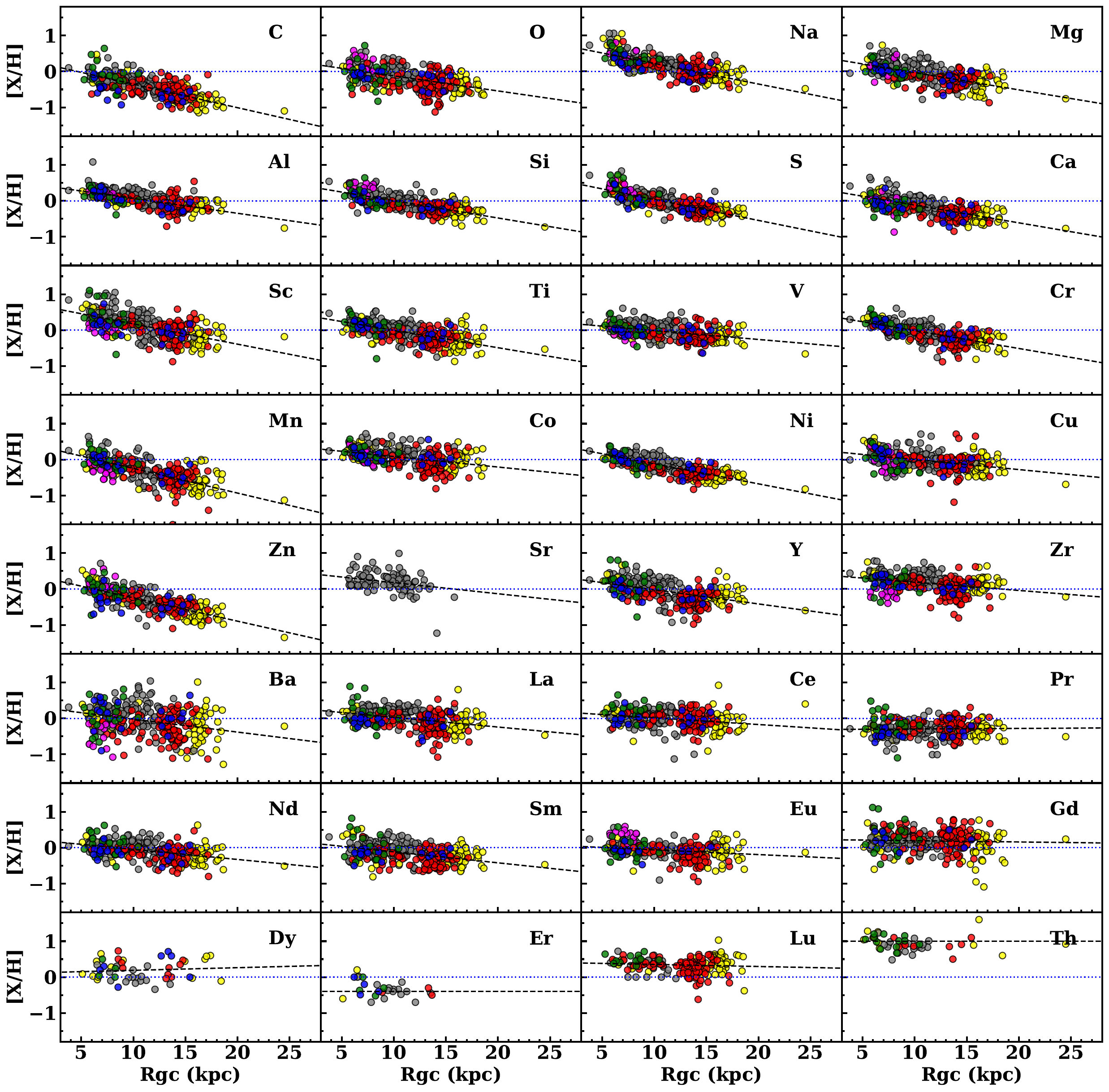}
    \caption{Radial galactic gradients for each element in the form of [X/H]. Grey points represent stars already published in \citet{Ripepi2021a,Trentin2023a,trentin2024}. Stars observed with UVES and published in \citet{Trentin2023a,trentin2024} are coloured in yellow and red, respectively. New stars presented in this paper are coloured in blue, magenta, and green for PEPSI, HARPS, and UVES observations, respectively (see also the legend in \ref{fig:iron_grad}).}
    \label{fig:gradients}
\end{figure*}

\section{Data analysis}\label{sec:spec}

\subsection{Stellar parameters}

The procedure and steps for both atmospheric parameters and chemical abundances estimation are the same as those reported in \citet{Ripepi2021a,Trentin2023a,trentin2024}.
The atmospheric parameters that need to be estimated are the effective temperature ($T_{eff}$), surface gravity ($\log g$), microturbulent velocity ($\xi$), and the joint effect of macroturbulence and rotational velocity, called the line-broadening parameter ($V_{broad}$). 
A diffused method used to estimate the effective temperature, regardless of the interstellar reddening, is the line depth ratio (LDR) method \citep{gray1991precise,kovtyukh2000precise}, with around 32 available LDRs per spectrum. 
To estimate $\xi$, we first measured the equivalent widths (EWs) of \ion{Fe}{I} with a semi-automatic {\tt Python} routine. As neutral Fe is insensitive to $\log g$, it is possible to construct atmospheric models, using ATLAS9 \citep{kurucz1993new}, by fixing $T_{eff}$ and $\log g$ and choosing a value of $\xi$. Thanks to the WIDTH9 code \citep{kurucz1981solar}, it is possible to convert the EWs to abundances. We extrapolated the value of $\xi$ for which the iron abundance is independent of EWs.
Lastly, $\log g$ was estimated by imposing that the iron abundance from \ion{Fe}{II} lines is the same as the one obtained with \ion{Fe}{I} lines. The list of 145 \ion{Fe}{I} and 24 \ion{Fe}{II} lines was extracted from \cite{Romaniello2008}. Errors for $\xi$ and $\log g$ were obtained by error propagation from the linear fits ([Fe/H] versus EWs and [\ion{Fe}{II}/H] versus $\log g$, respectively).
In Table~\ref{tab:logObservations} we report all the useful parameters used as inputs for the abundance analysis of the 136 spectra analysed in this paper and presented in the next section. 

\subsection{Abundances}\label{sec:abundances}

A spectral synthesis technique was used in order to avoid line-blending problems caused by line broadening. The technique is divided into three steps: i) generation of plane-parallel local thermodynamic equilibrium (LTE) atmospheric models using the ATLAS9 code; ii) synthesis of stellar spectra through SYNTHE \citep{kurucz1981solar}; iii) convolution of the synthetic spectra to take into account instrumental and line broadening. By matching the synthetic line to a selected set of observed metal lines, it is possible to evaluate this last parameter.
We divided the observed spectra into intervals of variable width (from 10 to 50 \AA, depending on the region of the spectrum) and performed a $\chi^2$ minimisation of the differences between observed and synthetic spectra to derive the chemical abundances. The algorithm, written in {\tt Python}, is based on multi-dimensional minimisation of a function using the downhill simplex method \citep{nelder1965simplex}.
Up to 32 different chemical species were estimated for each spectrum. 
The error on the single measurement was obtained by evaluating the uncertainties caused by varying the atmospheric parameters. For variations of $\delta T${\rm eff}$ = \pm 150$ K, $\delta \log g= \pm$\,0.2~dex, and $\delta\xi = \pm$\,0.3~km\,s$^{-1}$, we estimated a total contribution to the error of $\approx \pm$\,0.1~dex. When a single spectral line was detected for an element, we propagated the errors of the atmospheric parameters; otherwise, we averaged the abundance and propagated the errors.
We used the lists of spectral lines and atomic parameters from \citet{castelli2004spectroscopic}, which updated the original parameters from \citet{kurucz1995kurucz}. When necessary, we also checked the NIST database \citep{ralchenko2019nist}. Solar values used as a reference are from \citet{asplund2021}.
We evaluated the impact of non-LTE correction on our results using the grids published in \citet{amarsi20}, which explored a wide range of non-LTE corrections for several elements. In particular, according to the low metallicity of our targets, we neglected corrections for carbon, magnesium, silicon, calcium, and barium, while we applied a correction of 0.1 dex on our LTE sodium abundance.

\section{Results}\label{sec:results}

\subsection{Spectral lines detected and abundances}

In this section, we present the list of spectral lines through which it was possible to estimate up to 32 different chemical species in the final phase of the spectral analysis and discuss the estimated abundances. This list, shown in Table~\ref{tab:list_lines}, is updated from the one already presented in \citet{Ripepi2021a,Trentin2023a}. It is worth noting that for some elements (such as silicon, sulphur, scandium, cobalt, cerium, praseodymium, europium) the number of available lines is sensibly increased (if not doubled); for others (such as titanium and neodymium), several lines are available almost along the whole spectrum and we reported the most isolated and/or the strongest ones. In the last column of Table~\ref{tab:list_lines}, a brief note on each line was added, specifying whether it is a new line ('New') and if it was typically found partially blended with another line ('Blend' or 'New,Blend' to distinguish between the old and new lines, respectively). In the latter case, the estimated abundances were taken into consideration whenever the blending was not extreme and/or when no other line could be found for that element. 
Since metal-poor stars are also the farthest ones (see next section), not all the lines are always available, and their spectra are more affected by noise. On this basis, we expect that the increased number of available spectral lines might affect the results in the metal-poor regime.
An example is shown in Fig.\ref{fig:Zinc} for the Zinc case. While before only one weak line at $\lambda$\, 6362.338 {\AA}(right panel) was available for the spectra obtained with UVES, a stronger line at $\lambda$\, 4810.532 {\AA} (left panel) is now preferred. 
For these reasons, and to maintain the homogeneity in our analysis, we also re-estimated the abundances of the whole sample of stars previously published in the context of the C-MetaLL project, namely 290 stars, whose abundances have been published in \citet{trentin2024}.  
The final list of abundances for the entire dataset, i.e. including both the stars analysed in this work and those from our previous papers, is reported in Table~\ref{tab:abundances}. This is the baseline data that will be used in the following.

When comparing our previously published abundances with those obtained with the updated list we note that, for most of the elements (e.g. oxygen, carbon, vanadium, titanium, barium) the differences are minimal (below 0.1 dex); for others (scandium, copper, zinc, yttrium, zirconium), most of the metal-poorer stars (with [Fe/H]<-0.4 dex) present differences up to 0.2-0.4 dex. This is evident when comparing Fig.\ref{fig:ab_vs_iron}, where we plot the abundances (in their [X/Fe] form) versus the iron abundance (in the [Fe/H] form),  with Fig. 5 of \citet{trentin2024}. For these five elements, the discrepancy at low metallicities (yellow and red points, mostly observed with UVES) between the sample presented in \citet{Trentin2023a} and \citet{trentin2024} is reduced and less evident.
For the heaviest elements, i.e. from barium onwards, elements previously not found in the sample presented in \citet{Trentin2023a} were estimated (see the yellow points for cerium, samarium, europium, and gadolinium and the red points for samarium only).
Lastly, we detected for the first time four new chemical elements: dysprosium (Dy, atomic number 66), erbium (Er, atomic number 68), lutetium (Lu, atomic number 71), and thorium (Th, atomic number 90).

The trends of each element with respect to iron are similar to those previously found in \citet{trentin2024}. In more detail:
\begin{itemize}
    \item C, O: For these elements, we found a descending trend with increasing metallicity, with over-solar values for [Fe/H]<-0.5 dex and sub-solar values for [Fe/H]> 0.2 dex. In between, both elements are almost constant, with carbon at $\approx$-0.1 [C/Fe] dex and oxygen at solar values. Three new spectral lines were found for C (at $\lambda$\, = 6014.830 {\AA}) and oxygen (at $\lambda$\, = 5577.339 {\AA} and 6363.776 {\AA});
    \item Odd-Z elements (Na and Al): Sodium was found to be constant along the whole metallicity range, at around 0.4 dex, while aluminium decreases at a constant pace, with [X/Fe] $\approx$0 dex for solar metallicity.
    \item $\alpha$ elements (Mg, Si, S, Ca): All these elements show a decrease in their abundances with increasing iron, and flatten at [Fe/H] $\approx$ -0.5 dex. Two lines have been added for both magnesium and sulphur, while several new lines have been found for silicium and calcium (see Table~\ref{tab:list_lines});
    \item Scandium shows a peculiar behaviour, with a decreasing trend at low [Fe/H], a flattening at [Sc/Fe] $\approx$ 0.4 dex, and a hint of increasing again at higher metallicities ([Fe/H]>0.2 dex). Two new lines at $\lambda$\, = 5641.001 {\AA} and 5667.136 {\AA} were found, plus a strong blended line at $\lambda$\, = 5657.875 {\AA};
    \item Iron-peak elements (from Ti to Zn): Most of the iron-peak elements (titanium, vanadium, chromium, cobalt, and nickel) show a trend similar to the one described for the $\alpha$ elements, while manganese and zinc remain relatively constant for the whole iron range. Copper shows a different behaviour, with a flattened value ($\approx$0.4 dex at low metallicity and then starts decreasing at [Fe/H] $\approx$ -0.3 dex, reaching under-solar values. It is important to highlight, as was done for zinc, a new line found at $\lambda$\, = 5700.237 {\AA}, which is added to the other two lines (at $\lambda$\, = 5150.500 {\AA} and 5218.201 {\AA}) available in the spectral range of UVES, the second one generally heavily blended with an iron line $\lambda$\, = 5217.919 {\AA};
    \item Heavy elements (from Sr to Gd): An arch-form trend (already found in \citet{trentin2024}) is seen for zirconium and confirmed for barium (although with very scattered values), lanthanum, neodymium, and europium, and a descending trend for yttrium, cerium, praseodymium, samarium (also here with the presence of several outliers), and gadolinium. As was already mentioned, abundances for Ce, Sm, Eu, and Gd that were missing in the previous works have now been added;
    \item Dysprosium (Dy): Dysprosium is a heavy neutron capture element produced through the rapid neutron capture process (r-process). It is considered, together with europium and gadolinium, as a pure r-process element, since 98\%, 82\%, and 88\% of the solar abundances of Eu, Gd, and Dy are produced by the r process \citep{sneden2008}, respectively. As for the other r-process elements, their origin is still a matter of debate, with the most favoured hypothesis being the yields of neutron star mergers \citep{rosswog2018}, neutron star-black hole mergers \citep{fernandez2020}, fast-rotating supernovae with high magnetic fields \citep{cameron2003}, and hypernovae or collapsars \citep{brauer2021}. In the past, core-collapse supernovae (CCSNe)  had also been considered as a possible source, but recent simulations present limitations and do not provide the right conditions \citep[see e.g. ][and reference therein]{woosley1994,cowan2021}. The detected line, at $\lambda$\, = 5090.386 {\AA}, is highly blended with the FeI ($\lambda$\, = 5090.773 {\AA}) and appears for most stars as an asymmetry or a ‘bump’ in the shape of the latter. Nonetheless, this effect is for some objects strong enough for an estimation of the abundance to be made and, for higher values of [Dy/H], to distinguish the centre of the line. As is shown in the left panel of Fig.~\ref{fig:r_elements}, different values of the abundance radically change the deformation of the wing of the adjacent FeI line. Therefore, the estimated values of [Dy/H] cannot be trusted with certainty and need to be considered more as an upper limit. Its behaviour with respect to [Fe/H] reflects that of the other neutron capture elements, with abundances of the order of [Dy/Fe]$\approx$1 dex for [Fe/H]$\approx$ -0.5 dex and decreasing towards solar values at solar iron;
    \item Erbium (Er), lutetium (Lu), and thorium (Th): Abundance estimation of these elements had been made for the Sun and chemically peculiar stars of class A and B (Ap and Bp, respectively) \citep{cowley1987,leocat2004,lawler2008}. Chemically peculiar stars are main-sequence stars with strong magnetic fields and anomalies in their chemical compositions. The observed overabundances of these heavy and rare-earth elements in Ap and Bp stars, however, are not of Galactic origin but arise from radiative diffusion processes operating in their stable, magnetically structured atmospheres, leading to superficial and non-cosmogenic chemical anomalies. The spectral line we found for erbium is a very weak line at $\lambda$\, = 5028.905 {\AA}, which allowed it to be estimated for a few stars with good S/N (an example of the observed line is shown in the middle left panel of Fig.~\ref{fig:r_elements}). We detected for Lu a clear and isolated line at $\lambda$\, = 6221.891 {\AA}(middle right panel of Fig.~\ref{fig:r_elements}). The detected line for Th is at $\lambda$\,= 5989.045 {\AA}, highly blended with the close \ion{Nd}{II} line at $\lambda$\, = 5989.378 (right panel of Fig.~\ref{fig:r_elements}). As for Dy, the abundances for this element cannot be fully trusted.
    Like the other r-process elements, we found a decreasing trend with respect to [Fe/H], with [Er/Fe] ranging from 0.8 down to $-$0.8 dex, [Lu/Fe] ranging from 1.4 down to 0 dex and [Th/Fe] ranging from 0.5 to 1.5 dex. It is worth noticing how the green synthetic spectra in all the panels of Fig.~\ref{fig:r_elements} can be considered an indicative lower limit for the detection of Dy, Er, Lu, and Th, for spectra with S/N$\sim$80. At these abundances, the signal is almost indistinguishable from the noise, although we remind the readers that the trends themselves (in addition to individual abundance measurements) cannot be considered reliable with high confidence.
\end{itemize}

Lastly, we report in Table~\ref{tab:ab_smc} the estimated abundances for the two SMC stars Gaia DR2 4638315266636230400 and  Gaia DR2 4644478166748030976. We confirm the low-metallicity nature of these stars, with an estimated iron content of $-$0.75 dex and $-$0.89 dex, in agreement with typical values reported in the literature \cite[see e.g. ][ and reference therein]{Romaniello2008,Romaniello2022, breuval2024}. 

\begin{figure}
	\includegraphics[width=9cm]{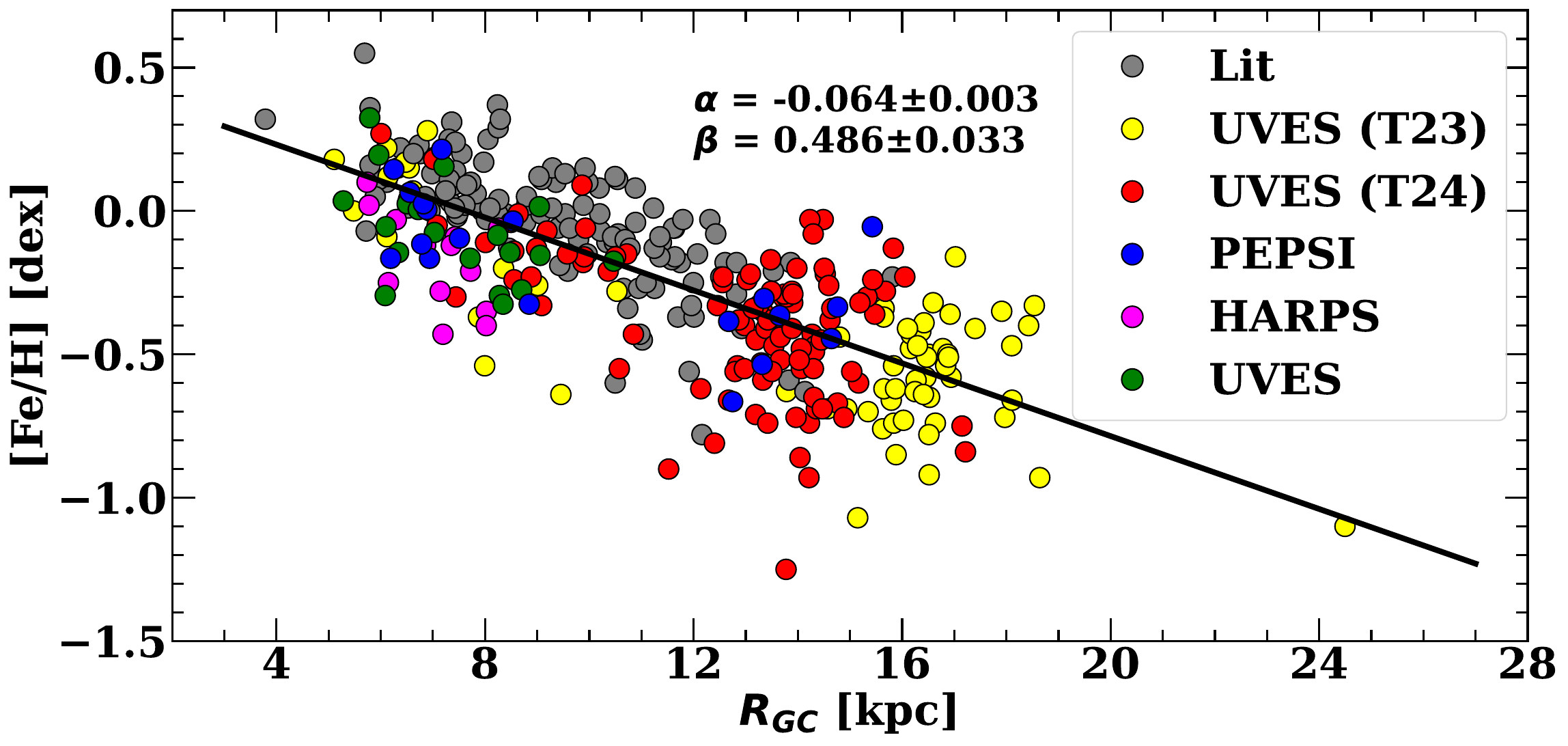}
    \caption{Galactic iron radial gradient (in the form of [Fe/H] = $\alpha\times R_{GC}+\beta$). The objects are colour-coded as in Fig.~\ref{fig:ab_vs_iron} and described in the legend. Grey points correspond to the DCEPs published in \citet{Ripepi2021a} and \citet{Trentin2023a} (except for those observed with UVES). The solid black line corresponds to the best-fitted line. Results of the fits are reported in both the figure and Table~\ref{tab:gradients_results}.}
    \label{fig:iron_grad}
\end{figure}

\begin{figure*}
	\includegraphics[width=17cm]{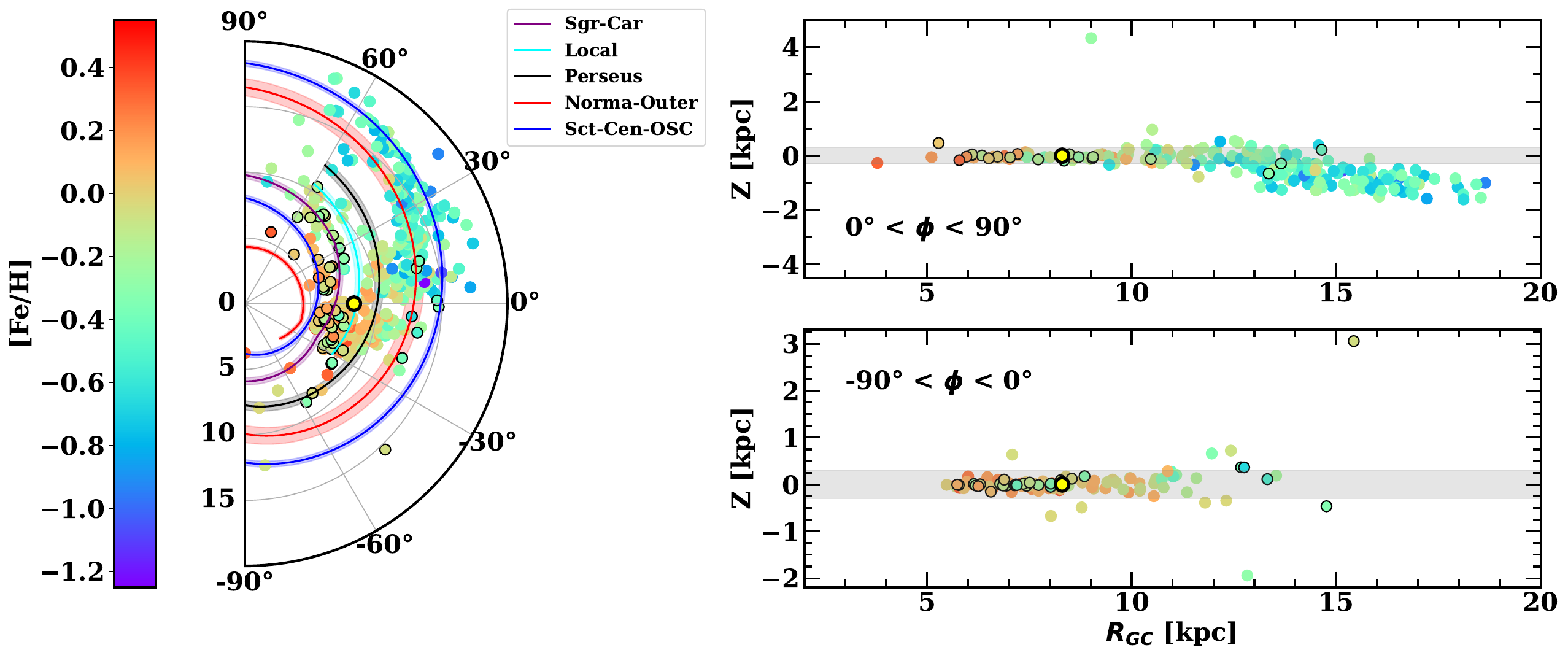}
    \caption{Left panel: Galactic polar distribution of the stars for azimuthal angle between -90 and 90 degrees. Circles with black borders highlight the targets presented in this work. All stars are colour-coded according to the measured iron abundance. The position of the sun is shown with a yellow-black symbol. Spiral arms adapted from \citet{Reid2019} are shown in different colours (see legend). Right panel: Cartesian representation of the height above or below the galactic plane as a function of $R_{GC}$ in two opposite directions with respect to the Galactic centre-Sun line (top and bottom panels, respectively). The colour coding and the symbols are as in the left panel. The grey area delimits the region at $|Z|<300$ pc}
    \label{fig:polar_zz}
\end{figure*}
\begin{figure*}
	\includegraphics[width=18cm]{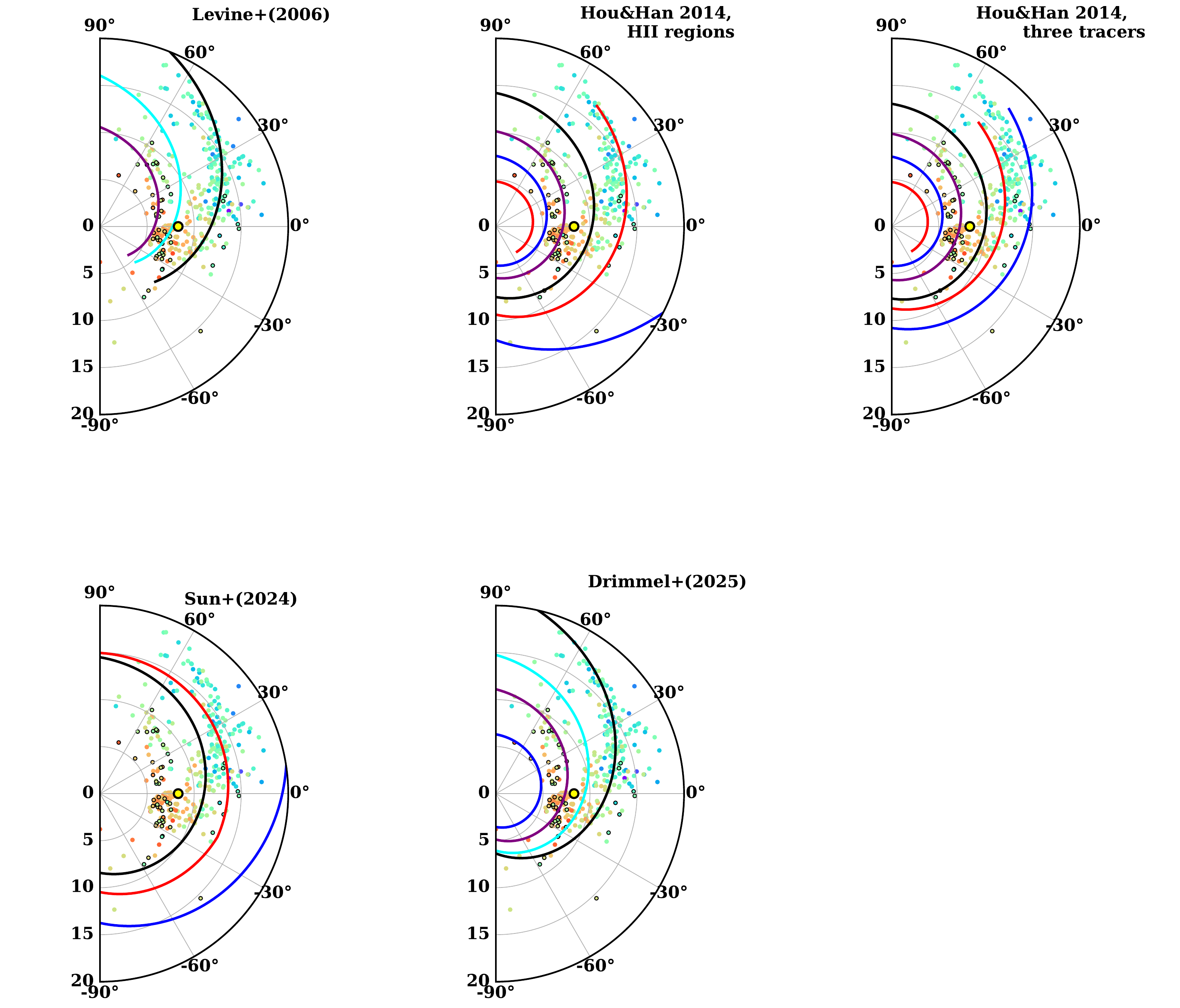}
    \caption{Spiral arm models from \citet{Levine_2006} (HI regions),\citet{Hou:2014} (using only HII regions or three different tracers), \citet{sun2024} (CO regions), and \citet{drimmel2025} (DCEPs) are superimposed on our targets in the top left, top middle, top right, bottom left and bottom middle panels, respectively. All the models (with the exception of \citet{sun2024}) are collected in the {\tt Python SpiralMap} package \citep{prusty2025}. Spiral arm colours are the same as those in Fig.~\ref{fig:polar_zz}.}
    \label{fig:arm_models}
\end{figure*}

\subsection{Radial gradients and spatial distribution}
Following the same procedure as in \citet{Ripepi2021a,Trentin2023a,trentin2024}, and described in Sect.~\ref{sec:observations}, we calculated the Galactocentric distance ($R_{GC}$) and the azimuthal angle of our targets and studied the radial gradients of all the measured chemical species (Fig.~\ref{fig:gradients}). The case of iron (usually used as a proxy of the metallicity) is shown  (in their [X/H] form, Fig.~\ref{fig:iron_grad}). For all the chemical species, we carried out a linear regression using the {\tt Python LtsFit} package \citep{Cappellari2013}, which implements a robust outliers removal, uses weights on both axes and estimates the error on the fitted parameters, together with the dispersion. The fitted slope for iron ($-$0.064 $\pm$0.002 dex $kpc^{-1}$) is in agreement with our latest results in \citet{trentin2024}, as well as most of the other chemical species. In more detail, light, $\alpha$ and iron-peak elements show a clear negative slope, with no precise trend as the atomic number increases. For neutron capture elements, they either have a shallower negative slope or gradients comparable with 0 (that is, no trend with $R_{GC}$). The apparent change in the abundance trend previously highlighted for Sc, Cu, Y, Zn, and Zr in \citet{trentin2024} is still present, but they now fall within the typical dispersion of the fit.
In Fig.~\ref{fig:polar_zz}, we focus our attention on the polar and vertical distribution of our star. As was done in \citet{trentin2024}, we superimposed the spiral arm models from \citet{Reid2019}. 
The upper right panel of figure 7 of \citet{Trentin2023a} shows three stars (OGLE GD-CEP-0120, OGLE-GD CEP-0123, and OGLE GD-CEP-0214) displaced from the others, at $R_{GC}\approx$18-19 kpc but $|Z|<500$ pc. We verified from the light curves that these stars are DCEPs and investigated their galactocentric distances calculated from the distances reported in \citet{Bailer2021}, and found a difference of around 4kpc, and no significant difference in $Z$. With the new estimated $R_{GC}$ ($\approx 14-15$ kpc), these stars are now distributed more coherently with the others. These values are the ones reported in Table~\ref{tab:programStars} and used in this work.
Three other stars, ATLAS J014.8936+45.4067, Gaia DR2 2246124001521354368, and V1253 Cen, caught our interest, since they are positioned at $Z$ = -1.9, 3.0 (visible in the bottom right panel of Fig.~\ref{fig:polar_zz} ), and 4.3 (top right panel of Fig.~\ref{fig:polar_zz}) kpc and distances from \citet{Bailer2021} agree with those estimated in our work within the errors. We studied the distribution of these objects in the energy-angular momentum diagram using \gaia~ proper motions and radial velocities and the {\tt Python} package {\tt galpy} \citep{bovy2015}\footnote{http://github.com/jobovy/galpy}. We found that these stars are located in the disc region of the diagram \citep[i.e. see the bottom right panel of figure 1 in][]{luongo2024}, and further analysis is needed.
Since the structure and identification of the spiral arms of the MW are still under debate, we used the  {\tt Python SpiralMap} package \citep{prusty2025} to discuss whether our targets can fit other models \citep[see ][ for detailed description of each of the models]{Levine_2006,Hou:2014,drimmel2025}. 
\citet{Levine_2006} proposed a logarithmic four-armed spiral model ($ln(R/R_0) = (\phi(R) -\phi_0 ) tan\psi$, with $\psi$ the pitch angle and $R_0$ and $\phi_0$ the Galactocentric distance and azimuth of the Sun, respectively) using a perturbed surface density map generated by the 21 cm hyperfine transition of neutral hydrogen. Three of these spirals, identified as the Carina, Local (or Orion), and Perseus arms, are plotted in the top left panel of Fig.~\ref{fig:arm_models}. A bigger pitch angle (of the order of 20-25°) characterises these arms, compared to those used by \citet{Reid2019} (typically $\approx$ 10° and always <20°). It is evident that stars closer to the centre of the MW are tracing the Carina arm, while our outermost targets nicely fit the Perseus one. 

\citet{Hou:2014} proposed a polynomial-logarithmic model ($ln (r) = a +b*\theta+c*\theta^2+d*\theta^3 $), using three different tracers (HII regions, Galactic molecular clouds, and methanol masers). In the top middle and top right panels of Fig.~\ref{fig:arm_models}, two different models are shown, whether only HII regions or all three tracers are used, respectively. In this case, six different arms have been identified: Norma, Scutum-Centaurus, Sagittarius-Carina, Perseus, Outer and Outer+1 Arms (OSC), and the Local arm (not shown Fig.~\ref{fig:arm_models}) that starts near the Perseus Arm, extends independently along the azimuth direction, and approaches the Carina Arm. The main difference with respect to the \citet{Levine_2006} model is the azimuth-dependence of the pitch angles, which results in a smaller ‘aperture’ of the arms. Indeed, in this case, our furthermost stars either fit the Outer arm or both the Outer and the OSC arms, although in the latter case a slight change of the fitted parameters would be needed to have a better fit (as was done in \citet{trentin2024} for \citet{Reid2019} models).

New models for the Perseus, Outer, and OSC arms were proposed by \citet{sun2024}, who analysed CO emission lines in more than 32,000 molecular clouds. These models, similar to what was done in \citet{Reid2019}, allow the pitch angle to abruptly change at a certain angle, called a ‘kink’. Their model is shown in the bottom left panel of Fig.~\ref{fig:arm_models}. In this case, only some of our furthest targets would trace the Outer arm \citep[as in][]{Reid2019}, but no spiral arm would fit those at $R_{GC}>15$ kpc.

More recently, \citet{drimmel2025} used new distances for almost 3000 Cepheids based on mid-IR WISE photometry \citep{skowron2025}, and identified four spiral arms (Scutum, Sagittarius-Carina, Local, and Perseus) described with a logarithmic model and shown in the bottom middle panel of Fig.~\ref{fig:arm_models}. Their models are in great agreement with those from \citet{Levine_2006}, and also in this case, our most distant stars better fit the Perseus arm.

\section{Discussion}\label{sec:discussion}

In the first part of this section, we compare our estimation of the radial metallicity gradient with those found in the recent literature.
When studying the radial metallicity gradient based on Cepheids, open clusters (OCs) are perfect targets for comparison. In most cases, information about the age is available, and it is possible to select clusters that are as young as Cepheids. In the fifth paper of the OCCAM project, \citet{carbajo2024} collected a total of 36 OCs with metallicities estimated from high-resolution red clump spectra. They found a gradient of $-$0.059$\pm$0.017 dex $kpc^{-1}$, in perfect agreement with ours. Moreover, they considered a second dataset (called OCCAM+) of 99 OCs composed of their sample complemented with OCs from the literature: Gaia-ESO Survey Data Release 5 \citep[GES DR5, ][]{Magrini2023}, Apache Point Observatory Galactic Evolution Experiment (APOGEE) DR17 \citep{myers2022}, and Galactic Archaeology with HERMES (GALAH) DR3 \citep{spina2021galah}. As in several other studies \citep{carrera2019,donor2020open,zhang2021a,myers2022}), they also detected a knee at around 11 kpc, over which the gradient flattens. In both cases of global slope ($-0.062\pm0.007$ dex $kpc^{-1}$) and inner slope ($-0.069\pm0.008$ dex $kpc^{-1}$), their results are in very good agreement with ours.

\citet{yang2025} studied the abundance gradients of Mg, Si, Fe, and Ni using a homogeneous sample of 299 OCs. Their sample combined astrometric and photometric data from \citet{hunt2023} (based on Gaia DR3) with metallicities from the Large Sky Area Multi-Object Fiber Spectroscopic Telescope (LAMOST) DR11 catalogue. These metallicities were determined using either the LAMOST Stellar Parameter pipeline \citep{luo2015} or the convolutional neural network (CNN) Stellar Parameter Convolutional Attention Network (SPCANet) \citep{wang2020}. Although their slopes are systematically smaller (in terms of the absolute value), for the two iron-peak elements there is agreement at the 1.5$\sigma$ and 1$\sigma$ levels for Fe and Ni, respectively.
It is worth mentioning that in both \citet{carbajo2024} and \citet{yang2025} it was found that the slope steepens with age, with the inner Galaxy young OCs showing lower [Fe/H] than old OCs, while in the outer Galaxy this trend reverses. For recent results regarding DCEPs, we refer the readers to the discussion in \citet{trentin2024,Trentin2023a} and section 6 of \citet{bono2024}.

The rest of this section is dedicated to the comparison of recent literature estimates of Dy, Er, Lu, and Th with those estimated for DCEPs.
In the context of the AMBRE project \citep{delaverny2013}, whose main goals are to provide stellar parameters and abundances through automated algorithms of ESO archived spectra and to create chemical and kinematical maps of Galactic stellar populations, Dy was estimated for 5479 stars in \citet{guiglion2018}. The observations are based on UVES, HARPS, GIRAFFE \citep[part of the FLAMES facility,][]{pasquini2002}, and the Fiber-fed Extended Range Optical Spectrograph (FEROS) \citep{stahl1999} instruments.
Through the study of [Eu/Ba], [Gd/Ba] and [Dy/Ba] as a function of metallicity (see their fig. 6), they found a clear indication of a diﬀerent nucleosynthesis history in the thick disc between Eu and Gd–Dy.
Measurements of Dy were carried out by \citet{mishenina2022}, based on 276 FGK dwarf stars in the galactic disc observed in \citet{mishenina2013} with the 1.93-m telescope at the Observatoire de Haute-Provence (OHP) and the echelle-type spectrograph Echelle spectrograph for Observations of Diffuse Interstellar bands and Exoplanets \citep[ELODIE,][]{baranne1996} and additional archival spectra collected with the Spectrographe pour l’Observation des Phénomènes des Intérieurs stellaires et des Exoplanètes \citep[SOPHIE][]{perruchot2008}. The trend of [Dy/Fe] against [Fe/H] for disc stars, shown in their figure 5 (which includes data from \citet{guiglion2018} and \citet{spina2018}), is in nice agreement with what is shown in Fig.~\ref{fig:ab_vs_iron}, with [Dy/Fe] ranging from  $\approx$ 0 and 1 dex for metallicities between -1 and 0 dex. Although for low metallicities (<$-$0.4 dex) our stars are typically more abundant in Dy ([Dy/Fe] between 0.5 and 1 dex) with respect to those presented in \citet{mishenina2022}, quantitative agreement is still found when comparing with data from \citet{guiglion2018}. The same is true for the other two ‘pure’ r-process elements analysed in this work, Eu and Gd.

In the context of the R-Process Alliance \citep[RPA, ][]{hansen2018}, whose main objective is to investigate the r-process through the study of r-process-enriched stars,  \citet{hansen2025}  observed the metal-poor star 2MASS J05383296–5904280 with the Space Telescope Imaging Spectrograph instrument mounted on the Hubble Space Telescope \citep[HST/STIS ][]{woodgate1998}. The metallicity of this star is very low, $-$2.59 dex, while Er, Lu, and Th are enhanced with abundances of [Er/H] = $-$1.38, [Lu/H] = $-$1.26 dex, and [Th/H] = $-$1.18 dex (that is, [Er/Fe] = 1.21 dex, [Lu/Fe] = 1.33 dex, and [Th/Fe] = 1.41 dex). These results agree with the behaviour found in this paper of enhanced r-process abundances for low-metallicity values. Dy, which was estimated as well ([Dy/H] = $-$1.24 dex, or equivalently [Dy/Fe] = 1.34 dex), is also enhanced and agrees with what was stated for the other elements. We point out that, as was already stated in Sect.~\ref{sec:results}, our measurements of Dy, Er, and Th are not definitive and their detection needs further attention.

\citet{Lyubimkov2022a} studied three red giant stars (EK Eri, OU And, and $\beta$ Gem, with [Fe/H] = 0.1 dex, $-$0.11 dex and 0.03 dex, respectively ) with strong magnetic fields using the NES echelle-spectrograph on the 6-m BTA telescope at the Special Astrophysical Observatory of the Russian Academy of Sciences. While these stars are thought to be descendants of magnetic Ap stars, no anomalies in the abundances of the heavy elements were found. In the case of erbium, for EK Eri and $\beta$ Gem abundances of [Er/H] of 0.04 and 0.0 dex were estimated, but no Lu or Th was detected. These values, as well as their Dy estimation, are coherent with our measurements at solar metallicities. In a subsequent study \citep{Lyubimkov2022b}, nine K-Giants with planets were analysed and a similar estimation of Dy and Er was obtained. More recently, the sample of three magnetic red giants presented in \citet{Lyubimkov2022a} was complemented with twenty other objects \citep{Lyubimkov2024}, but in this case only Dy was measured.

Other recent literature estimations of either Er, Lu, or Th in very metal-poor stars are available in \citet{saraf2023,lin2025,dasilva2025} and references therein. These stars all have metallicities of <$-$2.0 dex, and are considered among the oldest objects in the Universe, making them a useful tool in studying the chemical enrichment of the MW and its past interaction with other systems. In these cases, all three elements are enhanced, following the other heavy neutron-capture elements.
It is worth mentioning also the case of LAMOST J020623.21+494127.9, analysed in \citet{xie2024} and \citet{ashraf2025}. This star, although not very metal-poor (with a metallicity of [Fe/H] = $-$0.54 dex), is still an r-process-enhanced object  \citep[see table 2 in][]{xie2024} with a reported lutetium abundance of 0.20 dex. 

The only estimation of these three elements for Cepheids was reported for the first time in \citet{Kovtyukh2023} ( [Er/H] = +1.03 dex, [Lu/H] = +0.4 dex, and [Th/H] = +0.98 dex) for a DCEP in our sample (OGLE GD-CEP-1353, reported in this work as ASASSN-V J074354.86-323013.7), but not all the abundances match within the errors, probably due to the different code used by the authors (see their section 2).

\section{Summary}\label{sec:conclusion}

In this eighth work in the context of the C-MetaLL project, we collected and analysed a total of 136 spectra, corresponding to 60 DCEPs, 7 of which are repeated targets already published in previous C-MetaLL works, and 2 are SMC stars. One of the main characteristics of this dataset is that more than half of the objects are long-period Cepheids, with pulsational periods ranging from 15 to 70 days. Spectra were obtained with UVES@VLT, HARPS-N@TNG, and PEPSI@LBT. For each target, radial velocity, effective temperature, surface gravity, micro- and macro-turbulence velocities, and LTE chemical abundances were obtained. Following the instruction from \citet{amarsi20}, we applied a correction of 0.1 dex on our Na abundances to take into consideration non-LTE effects, while for the other elements, corrections were neglected. We confirm the low-metallicity nature of the SMC stars. 
We complemented our galactic sample with previously published targets in the context of the C-MetaLL project, for a final sample of 340 different stars (289 from past papers and 51 from present work, respectively). Using an updated list of trusted spectral lines, we revisited the abundances for all the stars for those elements with new available lines. While for most elements, such as O, C, V, Ti, Ba, and Nd, no statistical or systematic difference was detected between the previously estimated abundances and the new ones, the abundances of other chemical species (Sc, Cu, Zn, Y, and Zr) changed by more than 0.1 dex, especially in the low-metallicity regimee in which the available spectral lines are usually weaker and more affected by noise. The revisited abundances for these elements show a less pronounced or absent ‘jump’ in the radial direction.
Up to 33 chemical species have been detected, with Dy estimated for the first time in DCEPs and 3 elements (Er, Lu, and Th) estimated for several objects \citep[see ][ for the first detection of these elements in Cepheids]{Kovtyukh2023}. For all the elements we studied both their trends with respect to the iron content and their galactocentric radial gradients. In more detail:
\begin{itemize}
    \item For light elements, we found a decreasing trend with increasing metallicity, with [C/Fe] and [O/Fe] reaching sub-solar values and [Al/Fe] reaching solar values at higher [Fe/H]. Sodium instead is constant along the whole metallicity range, at [Na/Fe]$\approx$0.4 dex;
    \item $\alpha$ and iron peak elements show a decreasing behaviour followed by a flattening trend. Exceptions are given by: [Mn/Fe] and [Zn/Fe], which are relatively constant over the whole range; [Cu/Fe], with first a flattening trend and then a decrease at higher metallicities; and [Sc/Fe], which instead shows a slight enhancement for metal-rich stars;
    \item Neutron-capture elements, including the newly detected elements, show either an arch-form or a descending trend. Previously missing Ce, Sm, Eu, and Gd abundances from published DCEPs in the C-MetaLL project have now been estimated for several stars. The detected Dy line at $\lambda$ = 5090.386 {\AA} mostly appears as a deformation of the wing of a close \ion{Fe}{I} at $\lambda$ = 5090.773 {\AA}, but the effect is in some cases sensible enough to estimate this element. Er was estimated thanks to a weak line at $\lambda$ = 5028.905 {\AA}, while Lu for most stars presents a clear and isolated line at $\lambda$ = 6221.891 {\AA};
    \item We confirm a clear negative galactocentric radial gradient for most of our elements. Heavy neutron capture elements show instead shallower or null gradients, with [X/H] abundances mostly constant over a broad range of distances (5-20 kpc). The iron metallicity gradient $-0.064\pm0.002$ dex $kpc^{-1}$ is in good agreement with our previous work and recent literature gradients found for young OCs; 
    \item We studied the polar distribution of our stars in the MW disc, comparing different spiral arms models \citep{Levine_2006,Hou:2014,Reid2019,sun2024,drimmel2025}. Depending on which tracer and/or analytic function was used, different results were obtained. In this context, while the stars closest to the galactic centre nicely trace the Sgr-Car arm for all the models, those between 10 and 20 kpc can either trace the Perseus arm \citep{Levine_2006,drimmel2025}, the Norma-Outer arm \citep{Hou:2014,sun2024} or both the Outer and the OSC arms \citep{Hou:2014,Reid2019}.
\end{itemize}
It is important to stress that, beyond their importance in the study of the metallicity dependence of the DCEPs PL relations, the main focus of the C-MetaLL project, this work highlights how Galactic DCEPs can prove themselves as important tiles in tracing the young population of spiral arms and can be essential in the analysis of chemical evolution of the galactic disc. 
Most measurements of the heaviest elements are usually made for chemically peculiar main-sequence A and B stars, red giants with strong magnetic fields, probable descendants of Ap stars, and very metal-poor old stars ([Fe/H]<$-$2 dex). The detection of chemical species such as Er and Lu in young and relatively metal-rich stars such as DCEPs can constitute a turning point in the study of galactic enrichment, as well as offering a new scenario for investigating the production channel of neutron-capture elements and the differences between slow and rapid processes. However, it should be emphasised that, unlike in chemically peculiar main-sequence stars, where the large overabundances of heavy and rare-earth elements originate from atmospheric diffusion effects, the abundances measured in Cepheids genuinely reflect the chemical composition of the interstellar medium from which they formed, and thus provide reliable tracers of the Galactic chemical evolution.

\begin{acknowledgements}
Based on observations European Southern Observatory programs P105.20MX.001; P106.2129.001; 108.227Z; 109.231T; 110.23WM; 112.25NA, on the Telescopio Nazionale Galileo programmes A39TAC\_9; A40TAC\_11; A41TAC\_29; A42TAC\_15; A43TAC\_16; A44TAC\_27; A45TAC\_12; A46TAC\_15; A47TAC\_18. Based on observations obtained at the Canada-France-Hawaii Telescope (CFHT) which is operated by the National Research Council of Canada, the Institut National des Sciences de l´Univers of the Centre National de la Recherche Scientique of France, and the University of Hawaii. Based on observations obtained at the Large Binocular Telescope programs IT-2019B-014 and IT-2021-2022-24).
This research has made use of the
SIMBAD database operated at CDS, Strasbourg, France.
We acknowledge funding from INAF GO-GTO grant 2023 “C-MetaLL - Cepheid metallicity in the Leavitt law” (P.I. V. Ripepi). 
We acknowledge support from Project PRIN MUR 2022 (code 2022ARWP9C) "Early Formation and Evolution of Bulge and HalO (EFEBHO)" (P.I. M. Marconi), funded by European Union - Next Generation EU; INAF Large grant 2023 MOVIE (P.I. M. Marconi).
 This work has made use of data from the European Space Agency (ESA) mission Gaia (https://www.cosmos.esa.int/gaia), processed by the Gaia Data Processing and Analysis Consortium (DPAC, https://www.cosmos.esa.int/web/gaia/dpac/consortium). Funding for the DPAC has been provided by national institutions, in particular, the
institutions participating in the Gaia Multilateral Agreement.
This research was supported by the Munich Institute for Astro-, Particle and BioPhysics (MIAPbP), which is funded by the Deutsche Forschungsgemeinschaft (DFG, German Research Foundation) under Germany´s Excellence Strategy – EXC-2094 – 390783311.
This research was supported by the International Space Science Institute (ISSI) in Bern/Beijing through ISSI/ISSI-BJ International Team project ID \#24-603 - “EXPANDING Universe” (EXploiting Precision AstroNomical Distance INdicators in the Gaia Universe).
A.B.  thanks  the  funding  from  the  Anusandhan  National Research  Foundation  (ANRF)  under  the  Prime  Minister
Early Career Research Grant scheme (ANRF/ECRG/2024/000675/PMS).
\end{acknowledgements}

\bibliographystyle{aa} 
\bibliography{myBib} 
\onecolumn
\begin{appendix}
\section{Tables}
\begin{table}[htbp]
\footnotesize\setlength{\tabcolsep}{5pt}
\caption{Main properties of the 59 programme DCEPs}
\label{tab:programStars}
  \begin{tabular}{rcrrccc}
\hline
  \multicolumn{1}{c}{Star} &
  \multicolumn{1}{c}{Gaia\_source\_id} &
  \multicolumn{1}{c}{RA} &
  \multicolumn{1}{c}{Dec} &
  \multicolumn{1}{c}{Period} &
  \multicolumn{1}{c}{Mode} &
  \multicolumn{1}{c}{...} \\

  \multicolumn{1}{c}{} &
  \multicolumn{1}{c}{} &
  \multicolumn{1}{c}{deg} &
  \multicolumn{1}{c}{deg} &
  \multicolumn{1}{c}{days} &
  \multicolumn{1}{c}{} & 
  \multicolumn{1}{c}{}\\
  \hline
  2MASS J05580941+2802335 & 3431275314380598272 & 89.539167 & 28.04264 & 1.79323 & 1O & ...\\
  ASAS J115701-6218.7 & 5334675185024180864 & 179.253184 & -62.31222 & 26.45730 & F & ...\\
  ... & ... & ... & ... & ... & ... & ... \\
    \hline
\end{tabular}\\
\\
\\
\begin{tabular}{crrrrrrrr}
\hline
  \multicolumn{1}{c}{...} &
  \multicolumn{1}{c}{$G$} &
  \multicolumn{1}{c}{$G_{BP}$} &
  \multicolumn{1}{c}{$G_{RP}$} &
  \multicolumn{1}{c}{$R_{GC}$} &
  \multicolumn{1}{c}{$\phi$} &
  \multicolumn{1}{c}{$h$} &
  \multicolumn{1}{c}{Source} &
  \multicolumn{1}{c}{Notes}\\
  
  \multicolumn{1}{c}{} &
  \multicolumn{1}{c}{mag} &
  \multicolumn{1}{c}{mag} &
  \multicolumn{1}{c}{mag} &
  \multicolumn{1}{c}{kpc} &
  \multicolumn{1}{c}{deg} &
  \multicolumn{1}{c}{kpc} &
  \multicolumn{1}{c}{} &
  \multicolumn{1}{c}{}\\
  \hline
  ...& 12.432 & 12.953 & 11.734 & 14.6 $\pm$ 0.4 & 0.0165 &  0.21281 & PEPSI & SOS,SOS\\
  ...& 11.860 & 13.234 & 10.694 & 8.2 $\pm$ 0.4 & 0.92175 & -0.01348 & UVES & SOS,SOS\\
  ... & ... & ... & ... & ... & ... & ... & ... &...\\
\hline
\end{tabular}
\tablefoot{Meaning of the columns: Star = literature name of the DCEP. Gaia\_source\_id= $Gaia$ DR3 identifier. RA, Dec = equatorial coordinates at J2000. Period and Mode = period and mode of pulsation; F, 1O, and MULTI refer to fundamental, first overtone, and multimode DCEPs, respectively. $G$, $G_{BP}$, and $G_{RP}$ are the magnitudes in the $Gaia$ bands. $R_{GC},\phi$ and $h$ are the Galactic polar coordinates. Source shows the instrument used to observe the star. Notes report the origin of the periods (and modes) as well as of the photometry: ‘SOS’ means that the periods and the photometry come from $Gaia$ DR3 using the specific pipeline for the DCEP variables \citep[see][]{Ripepi2023}; ‘DR3’ means that the photometry is from the general DR3 source \citep[see][]{GaiaVallenari}, while ‘P21’ means that the periods are from      \citet{Piet2021}. 
A portion of the Table is shown here for guidance regarding its form and content. The machine-readable version of the full table will be published at the Centre de Données astronomiques de Strasbourg (CDS)}
\end{table}
\begin{table}[h]
\footnotesize\setlength{\tabcolsep}{3pt}
\caption{Observation information and atmospheric parameters for the 136 spectra analysed in this work.}
\label{tab:logObservations}
\begin{tabular}{rccrccccccr}
\hline
  \multicolumn{1}{c}{Star} &
  \multicolumn{1}{c}{MJD} &
  \multicolumn{1}{c}{Phase} &
  \multicolumn{1}{c}{Texp} &
  \multicolumn{1}{c}{S/N} &
  \multicolumn{1}{c}{$T_{eff}$} &
  \multicolumn{1}{c}{$\log g$} &
  \multicolumn{1}{c}{$\xi$} &
  \multicolumn{1}{c}{$V_{broad}$} &
  \multicolumn{1}{c}{$RV$} &
  \multicolumn{1}{c}{Source} \\

  \multicolumn{1}{c}{} &
  \multicolumn{1}{c}{days} &
  \multicolumn{1}{c}{} &
  \multicolumn{1}{c}{s} &
  \multicolumn{1}{c}{} &
  \multicolumn{1}{c}{K} &
  \multicolumn{1}{c}{dex} &
  \multicolumn{1}{c}{km {s$^{-1}$}} &
  \multicolumn{1}{c}{km {s$^{-1}$}} &
  \multicolumn{1}{c}{km {s$^{-1}$}} &
  \multicolumn{1}{c}{} \\

\hline
  2MASS J05580941+2802335 & 59475.94603 & 0.167 & 1500 & 187/290 & 6624 $\pm$ 27 & 1.58 $\pm$ 0.11 & 2.7 $\pm$ 0.4 & 15.0 $\pm$ 1.0 & -13.6 $\pm$ 0.1 & PEPSI\\
  ASAS J115701-6218.7 & 60396.53369 & 0.445 & 2100 &136 & 5253 $\pm$ 75 & 0.21 $\pm$ 0.01 &  3.2 $\pm$ 0.2 & 13 $\pm$ 1 & -20.5 $\pm$ 0.2 & UVES\\
  ... & ... & ... & ... & ... & ... & ... & ... & ... & ... & ... \\
\hline\end{tabular}
\tablefoot{The different columns report: the name of the star, the HJD at mid-observation, the phase at mid-observation, exposure time, S/N (when two arms are available, two values are reported for blue and red arm, respectively), effective temperature, logarithm of gravity, microturbulent velocity, broadening velocity, heliocentric radial velocity, and the instrument used to collect the HiRes spectroscopy. 
The phases were calculated adopting periods and epochs of maximum light from the $Gaia$ DR3 catalogue \citep{Ripepi2023}.
A portion is shown here for guidance regarding its form and content. The machine-readable version of the full table will be published at the Centre de Données astronomiques de Strasbourg (CDS).}
\end{table}
\begin{table}[h]
    \centering
    \caption{List of observed spectral lines in our spectra.} 
    \centering
    \begin{tabular}{lrrrrrrrrrr}
    \hline
    Ion & $\lambda$ ({\AA})  & $\log gf$& E$_i$~(eV)  & J$_i$  &  E$_f$(eV)~    & J$_f$  & $\log(\gamma_{r})$ & $\log(\gamma_{S})$ &  $\log(\gamma_{W})$ & Note \\ 
    \hline
    ... & ... & ... & ... & ... & ... & ... & ... & ... & ... & \\
\ion{Dy}{II} & 5090.386 & -2.332 & 0.10 & 7.5 & 2.54 & 7.5 & 0.00 & 0.00 & 0.00 & New,Blend\\
\ion{Er}{II} & 5028.905 & -2.067 & 0.05 & 5.5 & 2.52 & 6.5 & 0.00 & 0.00 & 0.00 & New\\
\ion{Lu}{II} & 6221.891 & -1.110 & 1.54 & 2.0 & 3.53 & 1.0 & 0.00 & 0.00 & 0.00 & New\\
\ion{Th}{II} & 5989.045 & -2.641 & 0.19 & 2.5 & 2.26 & 1.5 & 0.00 & 0.00 & 0.00 & New,Blend\\

    \hline
    \end{tabular}
    \tablefoot{Meaning of the columns: Ion = chemical element, $\lambda$ = central wavelength of the spectral line in angstrom units, Note = brief comment on the line: "New" = line not listed in previous C-MetaLL works; "N,B" = line not previously detected but typically found partially blended; "B" = line already presented in previous C-MetaLL works and typically found partially blended. A portion is shown here for guidance regarding its form and content. The machine-readable version of the full table will be published at the Centre de Données astronomiques de Strasbourg (CDS).}
    \label{tab:list_lines}
\end{table}

\begin{table}
    \footnotesize\setlength{\tabcolsep}{5pt}
    \caption{Estimated chemical abundances for the whole C-MetaLL dataset of 340 stars.}
    \label{tab:abundances}
    \begin{tabular}{cccccccccc}
    \hline
        \multicolumn{1}{c}{Star}&
        \multicolumn{1}{c}{[C/H]} & 
        \multicolumn{1}{c}{[O/H]} & 
        \multicolumn{1}{c}{[Na/H]} & 
        \multicolumn{1}{c}{[Mg/H]} & 
        \multicolumn{1}{c}{[Al/H]} & 
        \multicolumn{1}{c}{[Si/H]} & 
        \multicolumn{1}{c}{[S/H]} & 
        \multicolumn{1}{c}{...}\\
    \hline
    2MASS J05580941+2802335 & -0.67$\pm$0.13 & -0.25$\pm$0.09 & -0.25$\pm$0.01 & -0.33$\pm$0.03 & -0.31$\pm$0.11 & -0.23$\pm$0.18 & -0.28$\pm$0.05 &  ...\\
    ASAS-J115701-6218.7     & -0.43$\pm$0.18 & -0.36$\pm$0.09 & 0.25$\pm$0.13 & 0.08$\pm$0.13 & -0.01$\pm$0.03 & -0.02$\pm$0.11 & -0.09$\pm$0.06 & ...\\
    ... &... &... &... &... &... &... &... &... \\
    \hline
\end{tabular}\\
\\
\\
\begin{tabular}{ccccccccccc}
    \hline
        \multicolumn{1}{c}{...}&
        \multicolumn{1}{c}{[Ca/H]} & 
        \multicolumn{1}{c}{[Sc/H]} & 
        \multicolumn{1}{c}{[Ti/H]} &
        \multicolumn{1}{c}{[V/H]} & 
        \multicolumn{1}{c}{[Cr/H]} & 
        \multicolumn{1}{c}{[Mn/H]} & 
        \multicolumn{1}{c}{[Fe/H]} & 
        \multicolumn{1}{c}{[Co/H]} & 
        \multicolumn{1}{c}{[Ni/H]} & 
        \multicolumn{1}{c}{...}\\
    \hline
    ... & -0.6$\pm$0.11 & -0.28$\pm$0.11 & -0.24$\pm$0.09 & -0.64$\pm$0.2 & -0.33$\pm$0.16 & -0.61$\pm$0.12 & -0.44$\pm$0.11 & -0.07$\pm$0.15 & -0.41$\pm$0.13 & ...\\
    ... & -0.03$\pm$0.19& 0.15$\pm$0.19 & 0.05$\pm$0.15 & -0.1$\pm$0.1 & -0.08$\pm$0.13 & -0.21$\pm$0.12 & -0.09$\pm$0.12 & 0.04$\pm$0.11 & -0.12$\pm$0.10 &...\\
    ... &... &... &... &... &... &... &... &... &... &... \\
    \hline
    \\
    \hline
        \multicolumn{1}{c}{...}& 
        \multicolumn{1}{c}{[Cu/H]} & 
        \multicolumn{1}{c}{[Zn/H]} & 
        \multicolumn{1}{c}{[Sr/H]} & 
        \multicolumn{1}{c}{[Y/H]} & 
        \multicolumn{1}{c}{[Zr/H]} &
        \multicolumn{1}{c}{[Ba/H]} & 
        \multicolumn{1}{c}{[La/H]} & 
        \multicolumn{1}{c}{[Ce/H]} & 
        \multicolumn{1}{c}{[Pr/H]} & 
        \multicolumn{1}{c}{...}\\
    \hline
    ... & -0.28$\pm$0.1 & -0.65$\pm$0.1  & -- & -0.42$\pm$0.11 & -0.02$\pm$0.07 & 0$\pm$0.1 & -0.23$\pm$0.14 & -0.27$\pm$0.2 & -0.37$\pm$0.17 &...\\
    ... & -0.04$\pm$0.15 & -0.15$\pm$0.13 & -- & -0.04$\pm$0.12 & 0.06$\pm$0.08 & -0.07$\pm$0.2 & -0.22$\pm$0.19 & 0.09$\pm$0.1 & -0.41$\pm$0.17 & ..\\
    ... &... &... &... &... &... &... &... &... &... &... \\
    \hline
    \\
    \hline
        \multicolumn{1}{c}{...}& 
        \multicolumn{1}{c}{[Nd/H]} & 
        \multicolumn{1}{c}{[Sm/H]} & 
        \multicolumn{1}{c}{[Eu/H]} & 
        \multicolumn{1}{c}{[Gd/H]} &
        \multicolumn{1}{c}{[Dy/H]} &
        \multicolumn{1}{c}{[Er/H]} &
        \multicolumn{1}{c}{[Lu/H]} &
        \multicolumn{1}{c}{[Th/H]} &
        \multicolumn{1}{c}{Source}\\
    \hline
     ... &  -0.45$\pm$0.17 & -0.16$\pm$0.18 & -- & -- & -- & -- & -- & -- & PEPSI\\
    ... & -0.19$\pm$0.11 & -0.34$\pm$0.13 & -0.20$\pm$-0.19 & -0.08$\pm$0.2 & 0$\pm$0.1 & -0.52$\pm$0.10 & 0.45$\pm$0.10 & 0.84$\pm$0.1 & UVES\\    
    ... &... &... &... &... &... &... &... &... &... \\
    \hline
    \end{tabular}
    \tablefoot{The first and last columns report the name of the star and the source, respectively (see Table \ref{tab:logObservations} and \ref{tab:programStars}). The other columns report the estimated abundances (and relative errors) in solar terms for the chemical species analysed in this work. The machine-readable version of the full table will be published at the Centre de Données astronomiques de Strasbourg (CDS).}
\end{table}

\begin{table}
    \footnotesize\setlength{\tabcolsep}{5pt}
    \caption{Estimated chemical abundances for the SMC DCEPs Gaia DR2 4638315266636230400 and Gaia DR2 4644478166748030976.}
    \label{tab:ab_smc}
    \begin{tabular}{cccccccccc}
    \hline
        \multicolumn{1}{c}{Star}&
        \multicolumn{1}{c}{[C/H]} & 
        \multicolumn{1}{c}{[O/H]} & 
        \multicolumn{1}{c}{[Na/H]} & 
        \multicolumn{1}{c}{[Mg/H]} & 
        \multicolumn{1}{c}{[Al/H]} & 
        \multicolumn{1}{c}{[Si/H]} & 
        \multicolumn{1}{c}{[S/H]} & 
        \multicolumn{1}{c}{...}\\
    \hline
    Gaia DR2 4638315266636230400 & -0.50$\pm$0.13 & -0.45$\pm$0.05 & -0.88$\pm$0.08 & -0.57$\pm$0.17 & -0.66$\pm$0.10 & -0.70$\pm$0.10 & -0.53$\pm$0.15 &  ...\\
    Gaia DR2 4644478166748030976 & -0.72$\pm$0.11 & -0.7$\pm$0.1 & 0.76$\pm$0.11 & 0.87$\pm$0.19 & -0.57$\pm$0.08 & -0.48$\pm$0.11 & -0.73$\pm$0.14 & ...\\
    ... &... &... &... &... &... &... &... &... \\
    \hline
\end{tabular}\\
\\
\\
\begin{tabular}{ccccccccccc}
    \hline
        \multicolumn{1}{c}{...}&
        \multicolumn{1}{c}{[Ca/H]} & 
        \multicolumn{1}{c}{[Sc/H]} & 
        \multicolumn{1}{c}{[Ti/H]} &
        \multicolumn{1}{c}{[V/H]} & 
        \multicolumn{1}{c}{[Cr/H]} & 
        \multicolumn{1}{c}{[Mn/H]} & 
        \multicolumn{1}{c}{[Fe/H]} & 
        \multicolumn{1}{c}{[Co/H]} & 
        \multicolumn{1}{c}{[Ni/H]} & 
        \multicolumn{1}{c}{...}\\
    \hline
    ... & -0.70$\pm$0.16 & -0.38$\pm$0.10 & -0.32$\pm$0.10 & -0.46$\pm$0.17 & -0.52$\pm$0.14 & -0.71$\pm$0.15 & -0.76$\pm$0.16 & -0.36$\pm$0.11 & -0.73$\pm$0.11 & ...\\
    ... & -0.70$\pm$0.13 & 0.49$\pm$0.14 & -0.46$\pm$0.15 & -0.55$\pm$0.09 & -0.67$\pm$0.12 & -0.85$\pm$0.15 & -0.89$\pm$0.09 & -0.29$\pm$0.06 & -0.73$\pm$0.13 &...\\
    ... &... &... &... &... &... &... &... &... &... &... \\
    \hline
    \\
    \hline
        \multicolumn{1}{c}{...}& 
        \multicolumn{1}{c}{[Cu/H]} & 
        \multicolumn{1}{c}{[Zn/H]} & 
        \multicolumn{1}{c}{[Sr/H]} & 
        \multicolumn{1}{c}{[Y/H]} & 
        \multicolumn{1}{c}{[Zr/H]} &
        \multicolumn{1}{c}{[Ba/H]} & 
        \multicolumn{1}{c}{[La/H]} & 
        \multicolumn{1}{c}{[Ce/H]} & 
        \multicolumn{1}{c}{[Pr/H]} & 
        \multicolumn{1}{c}{...}\\
    \hline
    ... & -0.74$\pm$0.10 & -1.01$\pm$0.1 & -- & -0.56$\pm$0.14 & -0.44$\pm$0.16 & -0.11$\pm$0.09 & -0.34$\pm$0.11 & -0.17$\pm$0.16 & -0.38$\pm$0.20 &...\\
    ... & -0.79$\pm$0.02 & -0.8$\pm$0.08 & -- & -0.55$\pm$0.15 & -0.50$\pm$0.11 & -0.33$\pm$0.13 & -0.26$\pm$0.16 & -0.05$\pm$0.13 & -0.24$\pm$0.16 & ..\\
    ... &... &... &... &... &... &... &... &... &... &... \\
    \hline
    \\
    \hline
        \multicolumn{1}{c}{...}& 
        \multicolumn{1}{c}{[Nd/H]} & 
        \multicolumn{1}{c}{[Sm/H]} & 
        \multicolumn{1}{c}{[Eu/H]} & 
        \multicolumn{1}{c}{[Gd/H]} &
        \multicolumn{1}{c}{[Dy/H]} &
        \multicolumn{1}{c}{[Er/H]} &
        \multicolumn{1}{c}{[Lu/H]} &
        \multicolumn{1}{c}{[Th/H]} &
        \multicolumn{1}{c}{Source}\\
    \hline
     ... &  -0.19$\pm$0.12 & -0.28$\pm$0.11 & -0.16$\pm$0.12 & 0.03$\pm$0.12 & -- & 0$\pm$0.1 & 0.62$\pm$0.10 & 0.98$\pm$0.10 & UVES\\
    ... & -0.20$\pm$0.10 & -0.26$\pm$0.09 & -0.08$\pm$-0.03 & 0$\pm$0.13 & -- & -0.43$\pm$0.10 & 0.50$\pm$0.10 & 0.88$\pm$0.1 & UVES\\    
    ... &... &... &... &... &... &... &... &... &... \\
    \hline
    \end{tabular}
\end{table}

\begin{table}
    \centering
    \caption{Results of the fitting.} 
    \centering
    \begin{tabular}{lccc}
    \hline
    El & $\alpha$  & $\beta$ & rms \\ 
       & (dex kpc$^{-1}$) & (dex) & (dex)\\
    \hline
C  & -0.064 $\pm$ 0.003 &  0.280 $\pm$ 0.035 & 0.188\\ 
O  & -0.042 $\pm$ 0.004 &  0.293 $\pm$ 0.043 & 0.234\\ 
Na & -0.057 $\pm$ 0.003 &  0.795 $\pm$ 0.032 & 0.180\\ 
Mg & -0.048 $\pm$ 0.003 &  0.447 $\pm$ 0.037 & 0.210\\ 
Al & -0.041 $\pm$ 0.002 &  0.474 $\pm$ 0.025 & 0.143\\ 
Si & -0.048 $\pm$ 0.002 &  0.487 $\pm$ 0.026 & 0.158\\ 
S  & -0.056 $\pm$ 0.002 &  0.600 $\pm$ 0.026 & 0.148\\ 
Ca & -0.049 $\pm$ 0.002 &  0.370 $\pm$ 0.029 & 0.167\\ 
Sc & -0.056 $\pm$ 0.004 &  0.736 $\pm$ 0.043 & 0.261\\ 
Ti & -0.048 $\pm$ 0.003 &  0.487 $\pm$ 0.034 & 0.193\\ 
V  & -0.024 $\pm$ 0.003 &  0.231 $\pm$ 0.031 & 0.180\\ 
Cr & -0.049 $\pm$ 0.002 &  0.474 $\pm$ 0.029 & 0.167\\ 
Mn & -0.068 $\pm$ 0.003 &  0.439 $\pm$ 0.037 & 0.213\\ 
Fe & -0.064 $\pm$ 0.003 &  0.486 $\pm$ 0.033 & 0.10\\
Co & -0.029 $\pm$ 0.003 &  0.373 $\pm$ 0.040 & 0.216\\ 
Ni & -0.055 $\pm$ 0.002 &  0.439 $\pm$ 0.027 & 0.145\\ 
Cu & -0.027 $\pm$ 0.003 &  0.277 $\pm$ 0.035 & 0.213\\ 
Zn & -0.064 $\pm$ 0.003 &  0.400 $\pm$ 0.031 & 0.183\\ 
Sr & -0.032 $\pm$ 0.009 &  0.496 $\pm$ 0.088 & 0.198\\ 
Y  & -0.039 $\pm$ 0.003 &  0.368 $\pm$ 0.040 & 0.232\\ 
Zr & -0.023 $\pm$ 0.003 &  0.414 $\pm$ 0.035 & 0.202\\ 
Ba & -0.037 $\pm$ 0.006 &  0.347 $\pm$ 0.067 & 0.387\\ 
La & -0.027 $\pm$ 0.003 &  0.297 $\pm$ 0.039 & 0.218\\ 
Ce & -0.019 $\pm$ 0.003 &  0.203 $\pm$ 0.038 & 0.202\\ 
Pr &  0.001 $\pm$ 0.003 & -0.310 $\pm$ 0.040 & 0.209\\ 
Nd & -0.028 $\pm$ 0.003 &  0.235 $\pm$ 0.034 & 0.200\\ 
Sm & -0.030 $\pm$ 0.004 &  0.187 $\pm$ 0.044 & 0.238\\ 
Eu & -0.013 $\pm$ 0.003 &  0.077 $\pm$ 0.039 & 0.215\\ 
Gd & -0.004 $\pm$ 0.004 &  0.239 $\pm$ 0.052 & 0.272\\ 
Dy &  0.008 $\pm$ 0.011 &  0.103 $\pm$ 0.123 & 0.281\\ 
Er &  0.000 $\pm$ 0.014 & -0.400 $\pm$ 0.100 & 0.150\\ 
Lu & -0.005 $\pm$ 0.004 &  0.405 $\pm$ 0.057 & 0.197\\ 
Th &  0.000 $\pm$ 0.014 &  1.000 $\pm$ 0.800 & 0.150\\ 

    \hline
    \hline
    \end{tabular}
    \tablefoot{Coefficients of the linear fit of the form [X/H]=$\alpha\times R_{GC}+\beta$ with relative dispersion coefficient.}
    \label{tab:gradients_results}
\end{table}

\end{appendix}

\end{document}